\documentclass[twocolumn]{aastex6}
\usepackage{natbib}
\usepackage{rotating}
\usepackage{amsmath}
\usepackage{graphicx}
\usepackage{textcomp}
\usepackage{color}
\usepackage{tabularx}

\slugcomment{Draft version \today}
\shorttitle{Empirical Temperature Measurement in Circumstellar Disks.}
\shortauthors{Weaver E., Isella A., Boehler Y.}

\begin{document}

\title{Empirical Temperature Measurement in Protoplanetary Disks }

\author{Erik Weaver, Andrea Isella, Yann Boehler}
\affil{Department of Physics and Astronomy, Rice University\\ 6100
  Main Street, MS-108, Houston, Texas 77005}
\email{etw1@rice.edu}
 
\begin{abstract}
Accurate measurement of temperature in protoplanetary disks is critical to understanding many key features of disk evolution and 
planet formation, from disk chemistry and dynamics, to planetesimal formation. This paper explores the techniques available to 
determine temperatures from observations of single, optically thick molecular emission lines. Specific attention is given to issues 
such as inclusion of optically thin emission, problems resulting from continuum subtraction, and complications of real observations. 
Effort is also made to detail the exact nature and morphology of the region emitting a given line. To properly study and quantify these effects, this 
paper considers a range of disk models, from simple pedagogical models, to very detailed models including full radiative transfer. 
Finally, we show how use of the wrong methods can lead to potentially severe misinterpretations of data, leading to incorrect 
measurements of disk temperature profiles. We show that the best way to estimate the temperature of emitting gas is to analyze the
line peak emission map without subtracting continuum emission. Continuum subtraction,
which is commonly applied to observations of line emission, systematically leads to
underestimation of the gas temperature. We further show that once observational effects such
as beam dilution and noise are accounted for, the line brightness temperature derived
from the peak emission is reliably within 10-15\% of the physical temperature of the emitting
region, assuming optically thick emission. The methodology described in this paper will be 
applied in future works to constrain the temperature, and related physical quantities,
in protoplanetary disks observed with ALMA.
\end{abstract}


\section{Introduction}
\label{sec:intro}

Temperature is one of the key parameters describing the conditions
and dynamics inside protoplanetary disks. It controls disk chemistry,
and by setting the locations of various snow lines \citep{Qi2013}, it affects disk
structure. Thermal motion is a major component of gas dynamics \citep{BeSa1993}, and
must be fully understood in order to determine the extent of any nonthermal turbulent motion
present \citep{Flah2017,Teag2016}. Temperature also affects growth of dust grains, and by extension,
the growth of planetesimals \citep{Test2014,BlWu2008,WiCe2011}. Lastly, the temperature controls how
strongly and at which frequencies different components of the disk
emit.

Calculation of temperature in disks is nontrivial. Current methods
rely on radiative transfer codes, which use either Monte Carlo methods to
simulate billions of photons propagating through the disk to determine
a temperature at each point or use approximations to analytically solve 
the equation of radiative transfer. \citep[See ][]{ChGo1997, RaDC2006,Dull2007}. 
Energy is emitted primarily by the central
star, though contributions from external stellar radiation and, occasionally,
viscous heating are included. By modeling the dust and gas emission and comparing
to observation, the thermal structure of the disk can be estimated.
\citep[See, for example,][]{Rose2013,Piet2007}. Unfortunately, this process of
forward modeling is time consuming, computationally expensive, and
relies on assumptions, such as the vertical dust distribution of the
disk and dust opacity.

Temperatures can instead be inferred from optically thick molecular
emission lines. Spectral lines from molecules have long been powerful tools in 
radio astronomy, and in the current era of high-resolution ALMA observations, 
offer unprecedented information. The bulk of disk mass is in molecular hydrogen, 
which, lacking dipole moment, does not readily emit. However, other molecules, such 
as CO and CN are abundant in circumstellar disks, and emit strong spectral lines and 
can be used to trace the distribution of molecular hydrogen \citep{Guil2013}. Molecular 
lines such as CO rotational transitions are observable from ground based telescopes, 
such as ALMA. ALMA's ability to offer observations with sub-arcsecond angular resolution, 
allows disk structure to be spatially resolved. Additionally, because these lines can be optically 
thick, the spatial temperature distribution can be determined.

Observation of a source yields a map of the surface
brightness, which can be turned into a map of temperature via the
Planck equation. Brightness temperature can be calculated from maps of molecular
lines, for example, rotational transitions of CO. Other examples include NH$_{3}$, 
which can be used to measure temperatures of molecular cloud cores, and HCO$^+$. 
\citep[See][]{Juve2012,vnZd2001,Yen2016}. While the process of determining a brightness temperature can be quite simple, 
there are a number of subtleties that can affect the interpretations in terms of physical gas temperature. 
Under certain conditions, problems can arise due to subtraction of continuum emission, inclusion of 
optically thin emission, and complications from beam dilution. This paper will explore the conditions 
under which these problems arise, and what alternate techniques can be employed to deal with them.

Firstly, observations typically have the continuum emission subtracted
off in order to determine the pure line flux. The problem is that this 
process can end up removing line emission as
well as the continuum. This comes about because the continuum emission
is not a constant value across the spectrum, as normally assumed,
because the line absorbs the continuum. At the brightest points on
the line spectrum, the continuum emission is at its weakest, and for
strong lines, can be almost entirely attenuated. This means that if
the continuum emission is bright, much of the line emission can be
subtracted off, and any temperature measured will be too low, as will
any flux measured. The second problem is that no line is entirely
optically thick. While the line center may be bright and wholly optically
thick, the line wings quickly become optically thin, and when the
spectrum is integrated to produce the integrated emission map, this thin emission
is included, causing another underestimate. Lastly, the beam size and finite sensitivity of observations 
means that the ability to determine temperature is affected by both beam dilution, and 
noise. Beam dilution occurs when the beam size is larger than the spacial structure 
of the source, and leads to an underestimate, while noise can lead to both under and 
overestimates. These issues come about in various different 
situations, and can potentially have severe effects on analyses. The goal of our study is 
to quantify these effects for typical protoplanetary disks.

The structure of the paper is the following: Section 2 details a pedagogical model 
suitable for demonstrating the basic concepts and methods. Section 3 then repeats 
the analysis for a realistic disk model. Section 4 describes the results of simulated 
observation with ALMA, and conclusions and summary can be found in Section 5.

\section{Pedagogical model}

\subsection{Structure and emission}

In this section, we discuss the derivation of the temperature and density 
of a circumstellar disk based on a simple disk model with the following properties. 
We assume a disk of total mass M$_d$ distributed 
following a center-symmetric power law surface density 
\begin{equation}
\Sigma(r)=\Sigma_0\times \left( \frac{r}{r_0} \right)^{-p},
\end{equation}
between an inner radius ($r_{min}$) of 0.1~AU and an outer radius ($r_{max}$ )
of 100~AU.  $\Sigma_0$ is the surface density normalization at the reference radius $r_0$.
In this paper, $r_0$ is taken to be 10~AU, and the power law index, p, is taken to be 1.
$\Sigma_0$ is determined from the disk mass via 
\[
\Sigma_{0}=\frac{(2-p)\cdot M_d}{2\pi r_{0}^{p}\left(r_{max}^{p}-r_{min}^{p}\right)}
\]
The disk temperature distribution is also a radial power law with an index of -0.5, and is vertically
isothermal. Furthermore, the disk temperature does not depend on the disk mass. This is only an approximation, but it reasonably models
the interior of the disk \citep[see, e.g.][]{Dull2001} and, as discussed below, allows us to solve the radiative 
transfer equation for the dust and gas emission analytically. 
The temperature profile is 
\begin{equation}
T(r)=70~\textrm{K}\times \left(\frac{r}{r_0}\right)^{-0.5}
\end{equation}
This is chosen to roughly match the temperature profile resulting from a solar type star. 
The gas density $\rho(r,z)$ is calculated assuming that the disk is in 
hydrostatic equilibrium. Since the disk is vertically isothermal, 
the density is 
\begin{equation}
\rho(r,z)=\frac{\Sigma(r)}{\sqrt{2\pi}h(r)}exp\left(\frac{-z^{2}}{2h^{2}(r)}\right)
\end{equation}
and the pressure scale height is $h(r) = c_s(r)/\Omega(r) $, where  $c_s(r) = \sqrt{\gamma k_b T(r)/ \mu m_p}$ 
is the gas sound speed and  $\Omega(r) = \sqrt{G M_\star /r^3}$ is the Keplerian angular velocity. 
Using a mean molecular weight $\mu = 2.3$, an adiabatic index $\gamma=7/5$ appropriate for molecular 
hydrogen, a Boltzmann constant $k_b=1.38\times10^{-16}$ erg K$^{-1}$, a proton mass $m_p =1.67\times 10^{-24}$ g, 
a gravitational constant $G=6.67\times10^{-8}$~cm$^3$~g$^{-1}$ s$^{-2}$,
and a stellar mass $\textrm{M}_\star = 2\times10^{33}$~g, 
the pressure scale height is
\begin{equation}
h(r) = 0.5~\textrm{AU} \times \left( \frac{r}{r_0} \right)^{1.25}
\end{equation}

Neglecting scattering and assuming that the disk is in Local Thermodynamic Equilibrium, the emission from the disk is described by the radiative transfer equation 
\begin{equation}
\frac{dI_{\nu}}{ds}=K_\nu(s)(B_{\nu}(T)-I_{\nu}(s)),
\end{equation}
where $s$ is a spatial coordinate along the line of sight, and $K_\nu(s)$ is the total 
opacity, $K_\nu(s) = \kappa_\nu(s) \rho(s)$. 
The assumption of LTE is valid, provided that the number density is much greater than 
the critical density of the gas. In the case of protoplanetary disks, this is reasonable, given that the critical density, given by
\begin{equation}
n_{crit}=\frac{A_{ij}}{<\sigma\cdot v>},
\end{equation}
where $A_{ij}$ is the Einstein A Coefficient, $\sigma$ is the collisional cross section,
and $v$ is the velocity. The angle brackets indicate and average over the velocity distribution. 
This is simply the ratio of the radiative decay rate, given by the 
Einstein A value, to the rate coefficient for collisional de-excitation. The Einstein A 
values for CO can be found in databases, such as the Leiden Atomic and Molecular Database.
The rate coefficient for collisions between CO and various species, such as H$_{2}$ and 
atomic Hydrogen, can be found in databases such as the Basecol database \citep{Dube2012}. As an example, 
the Einstein A value for the $^{12}$CO J$=3-2$ transition is $6.910\times10^{-7}$ s$^{-1}$, while the rate coefficient 
for collisions between $^{12}$CO and H$_{2}$ is $5.995\times10^{-11}$ s$^{-1}$ cm$^{-3}$, at 40~K, though there is little 
variation with temperature. This gives a critical density of $1.15\times10^{4}$ cm$^{-3}$. This is a small number density 
compered to those of protoplanetary disks. For example, a disk with mass of 
$10^{-4}$ M$_{\odot}$ and an outer radius of 100~AU would have number densities in the 
midplane ranging from $10^{6} - 10^{8}$ cm$^{-3}$. The density 
drops quickly with altitude, but even for this light disk model, the density is  above critical for up to 4 scale heights 
at the inner radius and up to almost three at the outer radius. So, the critical density is smaller than the number 
density of the gas in all but the most remote and vacuous upper regions of the disk. 

If the disk is viewed face-on, the radiative transfer equation can be integrated 
analytically and the intensity as a function of the radius is (Appendix A)  
\begin{equation}
I_{\nu}(r)=B_{\nu}\left(T(r)\right)\left(1-\exp\left(-k_{\nu}(r)\cdot\Sigma(r)\right)\right)
\end{equation}
where $\kappa_{\nu}$ is the mass absorption coefficient. Because we are in LTE,
the source function is simply the Planck Function.
The optical depth of the disk emission is 
\begin{equation}
\tau_\nu(r)=\kappa_{\nu}(r)\cdot\Sigma(r)
\end{equation}

Here, $\kappa_{\nu}$ includes all the opacity sources. In the millimeter-wave  
regime, the disk opacity is dominated by dust and molecular gas,  $k_{\nu} = \kappa_\nu^d + \kappa_\nu^g$. 

Although the total opacity is simply the sum of the individual opacities, the dust and gas emission ($I_{\nu}^d$, $I_{\nu}^g$) can only be 
separated in the optically thin case, where equation 7 reduces to
\begin{equation}
I_{\nu}(r)=\underbrace{B_{\nu}(T(r))\kappa_\nu^d(r)\Sigma(r)}_{I_{\nu}^d}  + \underbrace{B_{\nu}(T(r)) \kappa_\nu^g(r)\Sigma(r)}_{I_{\nu}^g}
\end{equation}
If $\tau_{\nu} \geq 1$, the contributions from dust and gas each emit and attenuate each 
other, and so cannot be simply separated. The implications of this fact in understanding molecular line 
observations will be discussed below. 

In our model, we adopt dust opacities proper for spherical grains made of a mixture of  
silicates and carbonaceous material. The opacities are calculated following the procedure 
described in \cite{Isel2009}. In brief, the optical properties of the
different materials are combined using the Bruggeman theory with relative 
abundances as in \cite{Poll1994}. Opacities for dust grains with sizes $a$ 
between 0.01-1000~$\mu$m are calculated using Mie theory and 
averaged over a grain sized distribution $n(a)\propto a^{-3.5}$. 
The dust mass opacity at 0.87~mm is 1.44 cm$^2$ g$^{-1}$, which, for the disk model adopted here, 
gives an optical depth for 0.87~mm dust emission of $\tau_{0.87mm}(r) \sim 0.4 \times ({r/r_{0}})^{-1}$. 

Molecular gas absorbs and emits in lines, and its opacity depends on the 
temperature, density, and dynamics of the emitting material. In this section, we limit our analysis 
to rotational transition J$=3-2$ of $^{12}$CO ($\nu_0 = 345.796$ GHz), whose emission can be calculated assuming LTE.  
However, our results can be generalized to any line for which the LTE approximation holds true.
The molecular opacity assumes the following form:
\begin{equation}
\kappa_{ij}=\frac{c^{2}}{8\pi\nu_{0}^{2}}\cdot\frac{g_{j}}{g_{i}}n_{i}\phi_{\nu}A_{ji}\left(1-\exp\left(-\frac{h\nu_{0}}{kT}\right)\right)
\end{equation}
where $\nu_{0}$ is the rest frequency of the transition, $g_{i}$ is the multiplicity of the $i$th state,
 $n_{i}$ is the number density of $^{12}$CO particles in the $i$th state, $\phi_{\nu}$ is the line profile, $A_{ij}$ is the Einstein A value,
 and T the temperature. For a detailed derivation, see Appendix B.

\begin{figure*}[!t]
\centering
\includegraphics[width=0.48\linewidth]{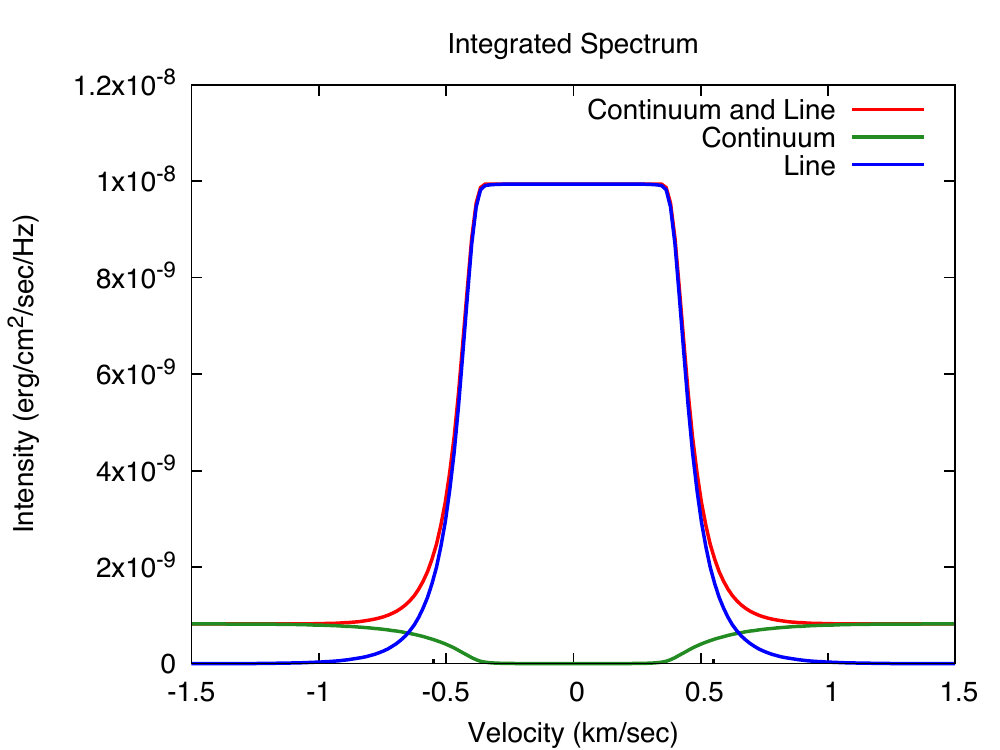}
\includegraphics[width=0.48\linewidth]{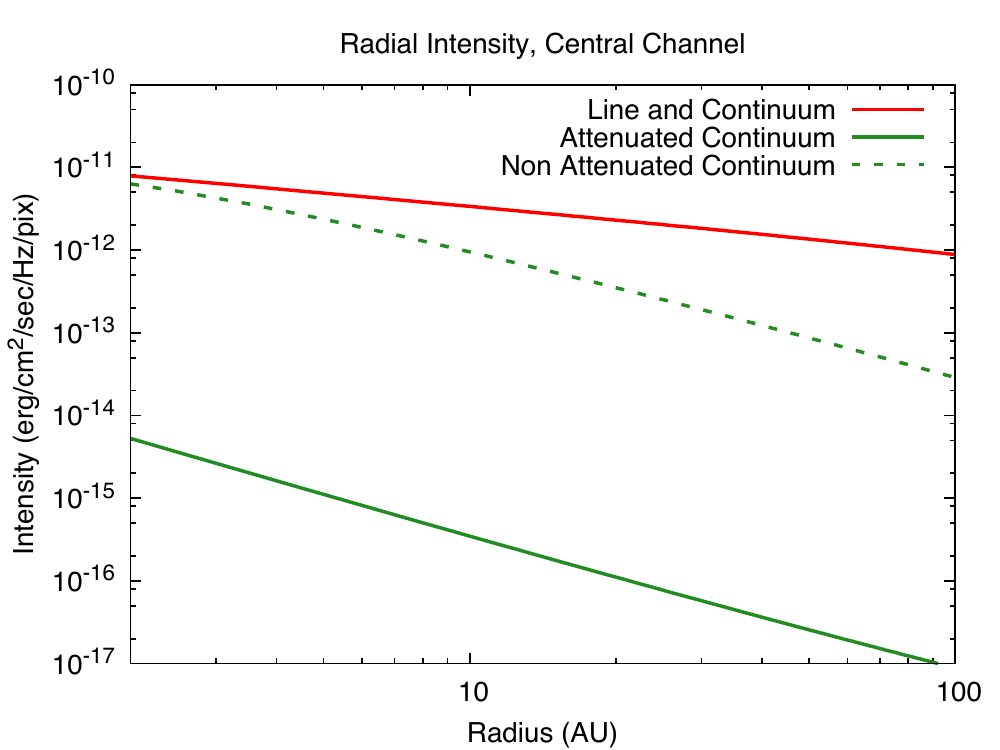}

\caption{\label{Fig:Spectrum} Left: Spatially integrated spectrum of the $^{12}$CO $J=3-2$ line, for a disk of mass 
$2\times10^{-2} M_{\odot}$, showing 
emission from both line and continuum (red), pure line emission (blue), and the actual continuum emission accounting for 
absorption by the line (green). At line center,the continuum emission is almost zero. The vertical lines at roughly 
±0.6km s$^{-1}$ indicate the boundaries between the optically thin and optically 
thick portions of the spectrum. Even for this relatively massive disk, a fair fraction of the spectrum 
is primarily optically thin. The square profile arises from the assumption a vertically isothermal temperature model, because 
all optically thick emission will have the same temperature. Right: Intensity as a function of radius at line center, for line and continuum (red),
attenuated continuum (green), and the continuum without line attenuation (dashed green). The attenuation lowers the continuum emission 
by three orders of magnitude.}
\end{figure*}

\begin{figure}[!t]
\centering
\includegraphics[width=\linewidth]{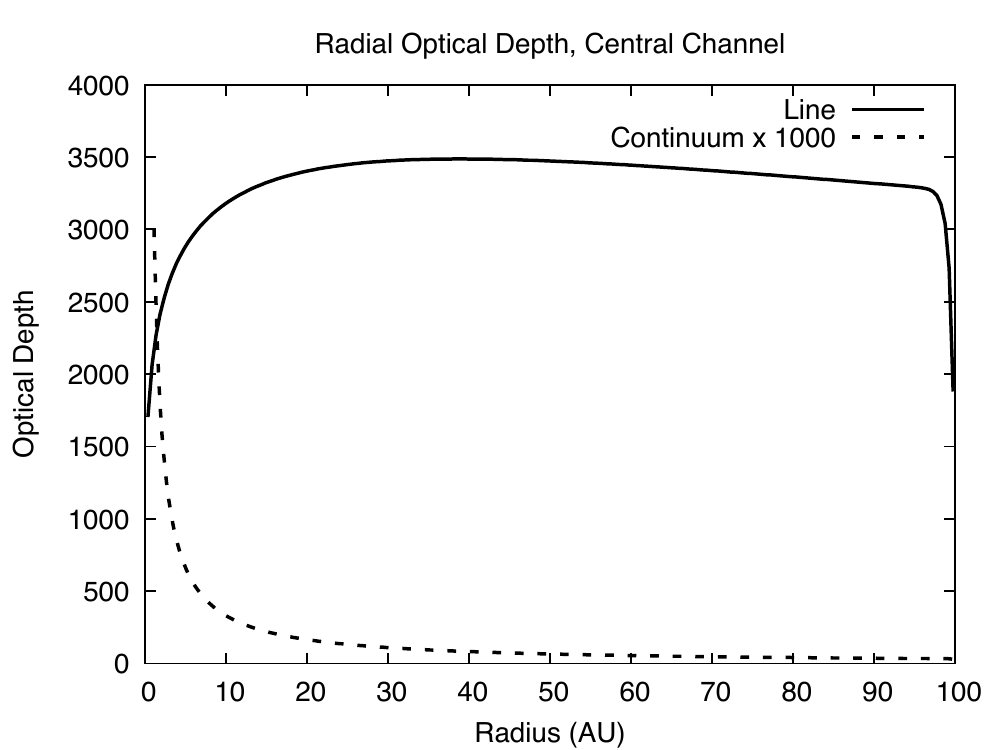}

\caption{\label{Fig:Tau} Optical depth as a function of radius, of the $^{12}$CO $J=3-2$ line, and continuum, for a disk of mass 
$2\times10^{-2} M_{\odot}$. The optical depth of the line (solid line) peaks around 35 AU. Further out in the disk, it decreases because the density 
drops quickly. Closer in, the optical depth also drops, because rising gas temperature depopulates the J=3 level. The dashed line shows the optical depth of the continuum, magnified by a factor of 1000, which behaves as a simple power law.}
\end{figure}

Figure~\ref{Fig:Spectrum} shows both the spatially integrated spectrum of  continuum and line emission, as well as 
their radial profile calculated at a frequency corresponding to the  line center, for a disk mass of $2\times10^{-2} M_{\odot}$. In this simulation, 
we adopted an H$_{2}/^{12}$CO abundance ratio of $10^5$ and an H$_{2}/$dust ratio of $10^2$. The integrated 
spectrum of the emission has a characteristic boxy shape, with a flat top, 
due to the fact that the disk model is vertically isothermal, and the central regions of the 
line in question are optically thick. Furthermore, because of the face-on inclination and because 
no turbulence is included in this model, the width of the line depends only on the thermal broadening 
(see Equation B12 and B13). As discussed above, the continuum and line emission cannot be easily separated in Equation 7. 
However, we can calculate the {\it pure} line emission $I_{\nu}^g$, i.e., the line emission attenuated by dust absorption, by  
subtracting the {\it pure} continuum emission $I_{\nu}^d$, i.e., the emission of the continuum but attenuated by the line, from the 
total emission given by Equation 7. The pure continuum and line emission are therefore respectively defined as
\begin{equation}
\label{Eq:faceOnSolnAdv}
I_{\nu}^d(r)=I_{\nu}(r) \frac{k_{\nu}^d(r)}{k_{\nu}^d(r)+k_\nu^g(r)} \,\,\,\, \text{and} \,\,\,\, I_{\nu}^g(r) = I_{\nu}(r) - I_{\nu}^d(r).
\end{equation}
Note that this assumes that the gas and dust are well mixed. This is an approximation, as the dust settles over time and 
millimeter-sized dust grains will concentrate towards the midplane.

The optical depth of the line is shown in Figure~\ref{Fig:Tau}, at line center. Although the temperature and density decrease 
monotonically with radius, the optical depth does not. This is because the opacity depends on the number density of CO molecules 
in the J=3 state, not just on the total number of molecules. In the inner disk, the temperature is very high, and the CO molecules 
equilibrate between a large number of level, so the third is less occupied. Further out in the disk, the dropping temperature forces more 
molecules into the J=3 state. The maximum population in the J=3 state occurs at about 30K. On the other hand, past a certain point, 
the dropping temperature and total CO number density eventually 
becomes the dominant effect, and the optical depth decreases again. The peak occurs at around 35 AU, where the line has an optical 
depth of $3.5\times10^{3}$, while the dust optical depth peaks at the inner radius, with a value of just 3. CO molecules therefore 
absorb practically all the continuum 
emission, causing it to drop by about three orders of magnitude compared to the continuum intensity outside the line. 
Similarly, at the outer radius of the disk, the optical depth of the line emission exceeds that of the continuum by 
5 orders of magnitude. In practice, since the optical depth of the CO emission close to the line center is very large, 
no dust emission can escape from the disk. The part of the spectrum where the line emission has an 
optical depth greater than unity is indicated in the left panel of Figure~\ref{Fig:Spectrum}, by the region between the two vertical lines.  
As discussed below, the absorption of continuum emission from CO molecules can have a large effect on the derivation 
of the gas temperature (and density) from the continuum-subtracted line intensity.

\subsection{Derivation of the disk temperature}

For thermal radiation, the intensity of the emission depends on temperature of the 
emitting material. In radio astronomy, it is a common procedure to measure intensities in units 
of Brightness Temperature, defined as the temperature of a blackbody emitting the same 
power per unit of solid angle per unit of bandwidth.  Intensities can therefore be converted 
into Brightness Temperature by inverting the Planck Function
\begin{equation}
T_{B} = \frac{h\nu}{k_{B}}\left(log\left(\frac{2h\nu^{3}}{c^{2}I_{\nu}}\right)+1\right)^{-1}
\end{equation}
If the emission is optically thick, the intensity depends only on the temperature of the emitting material
($I_{\nu}= B_{\nu}(T)$, Equation 7), and the brightness temperature is  
therefore equal to the physical  temperature ($T_B = T$). 
In general, the brightness temperature is related to the physical temperature by the 
relation 
\begin{equation}
B_\nu(T_B)=B_\nu(T)\left(1-e^{-\tau_\nu}\right)
\end{equation}
where $\tau_\nu$ is the optical depth of the emission (Equation 8) \footnote{In radio astronomy, it is 
common to defined the brightness temperature assuming Rayleigh-Jeans approximations. In this case, 
Equation 13 simplifies to $T_B = T(1-e^{-\tau_\nu})$.  However, since the Rayleigh-Jeans approximations
might not hold in the cold midplane regions of  protoplanetary disks, we prefer to define the brightness 
temperature using the Planck function.}
In reality however, a telescope like ALMA does not measure the intensity of the emission $I_{\nu}$ but its flux density $S_{\nu}$, 
which is the intensity integrated over the solid angle corresponding to the 
angular resolution of the telescope \citep[see][for a comprehensive review of this topic]{CoRa2016}. 
Flux densities are typically expressed in units of Jy beam$^{-1}$, where the beam is  
the full width half maximum of the synthesized clean beam. In the context of the synthetic 
models discussed in this Section, the relevant solid angle to calculate the flux density is that 
subtended by the pixel size of our simulated images.  The brightness temperature 
can still be defined using Equation 12, where the intensity is derived from the synthetic (or measured)
flux density using the relation
\begin{equation}
I_{\nu} = S_{\nu} / \theta^2, 
\end{equation} 
where $\theta$ is the angular size of the pixels of synthetic model (or the size of the synthesized  
beam in the case of real observations). This relation implies that the brightness temperature 
derived by our models (or by ALMA observations) is the average of the local brightness temperature calculated 
across the pixel (or the ALMA beam). 
As a consequence, if the emission of a thermal source is optically thick but arises from a region with 
angular area $A$ smaller then the area of the beam $\pi \theta^2$, its brightness temperature will be lower than its physical temperature by an 
amount equal to the ratio between the source size and the beam area. This effect takes the name 
of {\it beam dilution} and, as discussed in Section 4, has a direct impact on the observations of  
molecular line emission from protoplanetary disks. Conversely, if the emitting area is larger than the pixel (or beam), 
the measured brightness temperature is not affected by beam dilution. In summary, the relation between brightness temperature and physical 
temperature is 
\begin{equation}
B_\nu(T_B)= 
\begin{cases} 
B_\nu(T)\left(1-e^{-\tau_\nu}\right) \frac{\pi \theta^2}{A}, & \text{if } A < \pi \theta^2 \\
B_\nu(T)\left(1-e^{-\tau_\nu}\right) , & \text{if } A \geq \pi \theta^2 
\end{cases}
\end{equation}

If the emission is optically thick and spatially resolved, as for example in the case of ALMA observation of 
CO line emission from nearby protoplanetary disks \citep[see, for example,][]{Schw2016}, the exact knowledge of the optical depth is not 
necessary to derive the gas temperature. However, since the observers rarely know the exact optical depth, 
and because real disks are not vertically isothermal, the measured brightness temperature does not give 
the temperature at known points in the disk, but rather describes a general region. Lines of differing optical 
depth can therefore be used to probe different regions of the disk to determine the vertical 
temperature structure. (See, for example, \cite{Dart2003}, \cite{Isel2007}, or \cite{Rose2013}). The case 
of non vertically isothermal disks is discussed in Section 3.  

\subsubsection{Effect of Continuum Subtraction}

Temperatures of protoplanetary disks can be determined from measuring the flux 
density of optically thick lines (if the area of the emitting source is known), but since an observation combines emission from both the
line and the continuum, the contribution of the continuum must be quantified. 
The simplest and most common procedure is to assume that
across the spectral width of the line, the continuum emission is approximately
flat, and subtract it off. However, this procedure, commonly known as {\it continuum subtraction}
can lead to the removal of line emission if, as discussed in the previous section and shown in Figure~\ref{Fig:Spectrum},  
the line is optically thick and absorbs part or all of the underlying continuum. 
This effect is most severe in the denser regions of protoplanetary disks where both line and continuum emission 
tend to be optically thick. Standard continuum subtraction is accurate only in the optically 
thin limit represented by Equation 9. 

The effects of continuum absorption on line emission are often considered in astrophysical applications, 
but here it is important to include the effects of line absorption on the continuum. Because an optically thick line 
can almost completely attenuate continuum emission, the emission at the peak of the line spectrum 
is entirely that of the line. The amount of line emission removed by assuming a flat continuum 
increases as the optical depth
of the dust continuum emission increases, and it is maximized when both continuum and line emission are 
optically thick. 

Because of this, the effect is not constant
through the disk. The outer region has thinner dust, and a lower continuum
emission level. In the inner disk, however, the effect is particularly
severe, as the dust is much brighter where both the temperature and
the density are higher.  In our vertically isothermal model, all optically thick emission will have the same magnitude, so if the 
dust reaches optical depth greater than one, continuum subtraction will remove all
emission. In more physically realistic temperature models, the effect
may not be so severe as to remove all emission, but may still lead to an 
erroneous estimate of the true line emission. 

The (sub)-millimeter wave dust continuum emission has been found to be optically thick in 
several protoplanetary disks, including HL Tau, HD~163296, HD~142527, and IRS~48 \citep{Jin2016, Isel2016, Casa2013, Boe2017a, vdMa2016}.
Integrated intensity maps of the continuum-subtracted molecular line emission from these sources show that 
the line emission is strongly suppressed where the dust continuum emission is more intense. In particular, 
inner cavities has been observed in the $^{13}$CO and C$^{18}$O emission toward the center of the  HD163296 disk, 
where the dust continuum is optically thick. Similarly, a lack of molecular line emission has been observed in the optically 
thick dust crescents around IRS~48 and HD~142527 \citep{vdPl2014}. These features can be explained by line removal 
by continuum subtraction.

\begin{figure*}[!t]
\centering{}
\includegraphics[width=\linewidth]{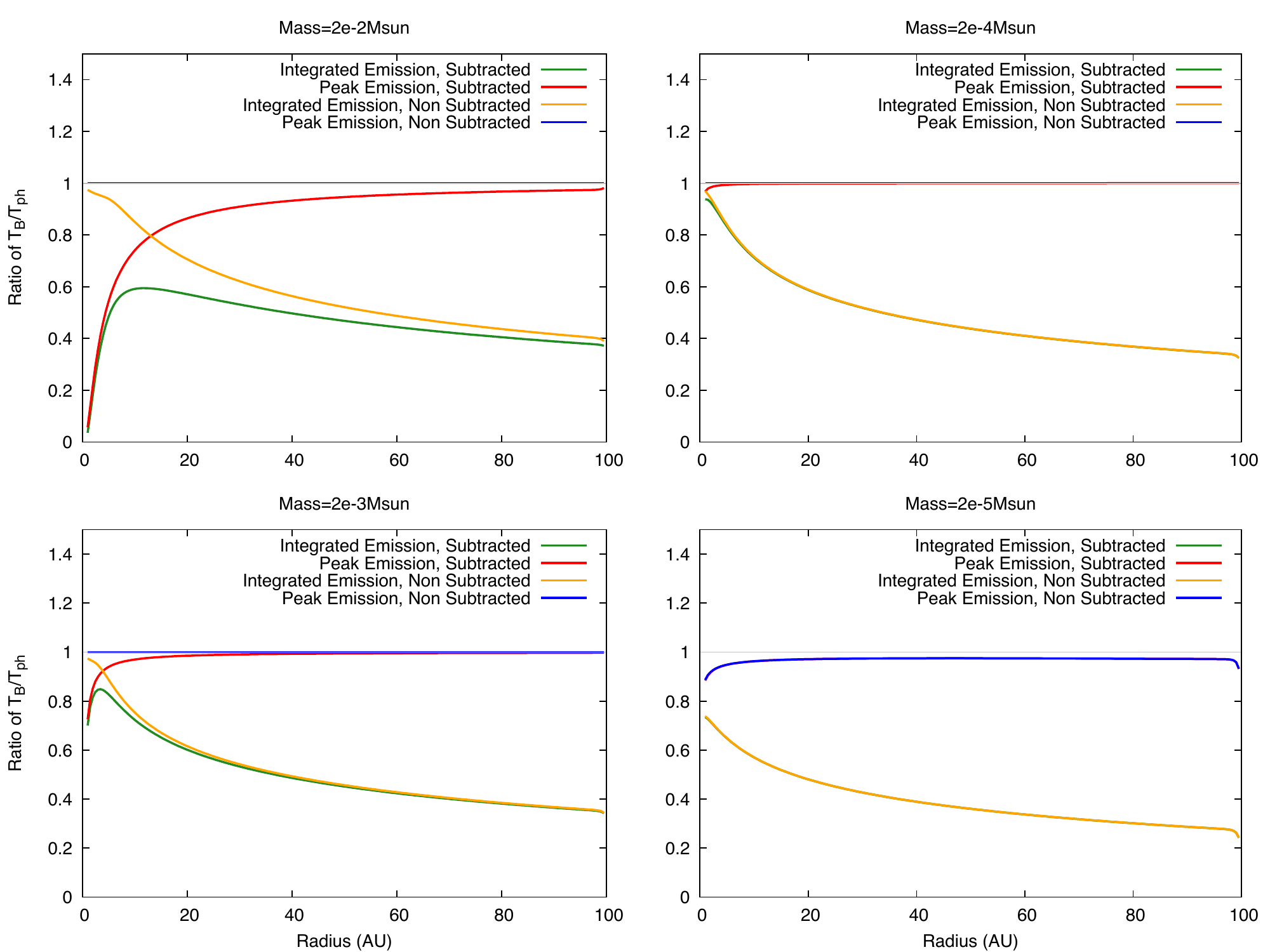}
\caption{\label{Fig:Isothermal} Ratios of brightness temperature to physical temperature for a 
variety of different methods and models. Each panel corresponds to a 
different disk mass, ranging from $2\times10^{-2} M_\odot$ to $2\times10^{-5} M_\odot$. 
Within each panel, the four lines correspond to integrated emission map and peak emission Map 
methods, both with and without continuum subtraction.
}
\end{figure*}

Figure~\ref{Fig:Isothermal} shows the effect of continuum subtraction on the $^{12}$CO $J=3-2$ arising from  
disks with masses of $2 \times 10^{-2}$, $2 \times 10^{-3}$, $2 \times 10^{-4}$, and 
$2 \times 10^{-5}$ M$_\odot$. This range covers the range of masses of nearby protoplanetary disks.
We adopted the pedagogical disk structure discussed above and calculated the brightness temperature $T_B$ of the 
line both before and after continuum subtraction. Furthermore, for each case we calculated the brightness temperature 
from both the spectrally integrated  (Moment 0) map and the peak emission map. The integrated emission map is calculated by 
averaging over the full spectral extent of the line (typically around ±25km s$^{-1}$ for the inclined disk models), 
while the peak emission is the emission recorded at the peak of the spectral line. 
Because the chosen line is mostly optically thick and the synthetic model is 
not affected by beam dilution, the brightness temperature of the line should be very close to the physical temperature $T$
of the disk, and the ratio $T_B/T$ should therefore be close to 1. 
We find instead that the brightness temperature derived from the continuum subtracted line emission can be much 
lower than the physical temperature. The effect of continuum subtraction is most evident in the most massive 
disk model, where the brightness temperature of continuum-subtracted lines (red and green curves) sharply drops to zero within 
about 10 AU from the center. In practice, both the peak emission and integrated intensity maps 
(red and green lines) show an artificial inner cavity for the heaviest model when 
the continuum is subtracted. Obviously, the brightness temperature derived from the 
integrated intensity is always lower than the physical temperature. This is because the integrated line 
emission includes optically thin contribution from the line wings. However, as will be explained in Section 4, 
there are definite advantages to the use of integrated emission maps in their robustness to the effects of both noise 
and beam dilution.

The effect of continuum subtraction diminishes with the intensity of the continuum, and therefore with the 
mass of the disk, and becomes negligible in disks with masses less than about $10^{-4}$ M$_\odot$. 
At disk masses lower than $10^{-5}$ M$_\odot$, the optical depth at the peak of the $^{12}$CO 3-2 line drops below 5 and 
the brightness temperature of the line is systematically lower than the physical disk temperature. Interestingly, the 
optical depth of the line has a minimum in the innermost (and densest) regions of the disk. This counterintuitive 
behavior is because the level populations of the CO rotational modes depend on the temperature, and in the inner disk, 
the very high temperature depopulates the third level. This leads to a lower optical depth for the 3-2 line, despite the 
larger gas density.

For weaker lines, where the optical depth of the line is not significantly greater than
that of the dust, non continuum-subtracted peak emission maps will
still offer better results than continuum-subtracted integrated emission maps,
but the brightness temperature will still fall short of the physical temperature if the optical depth is low. 
Nonetheless, for all models, the non-continuum subtracted peak emission map 
(orange line) gives the closest result to the physical temperature.

\subsection{Effect of Gas to Dust Ratio}

\begin{figure*}[!t]
\centering{}
\includegraphics[width=\linewidth]{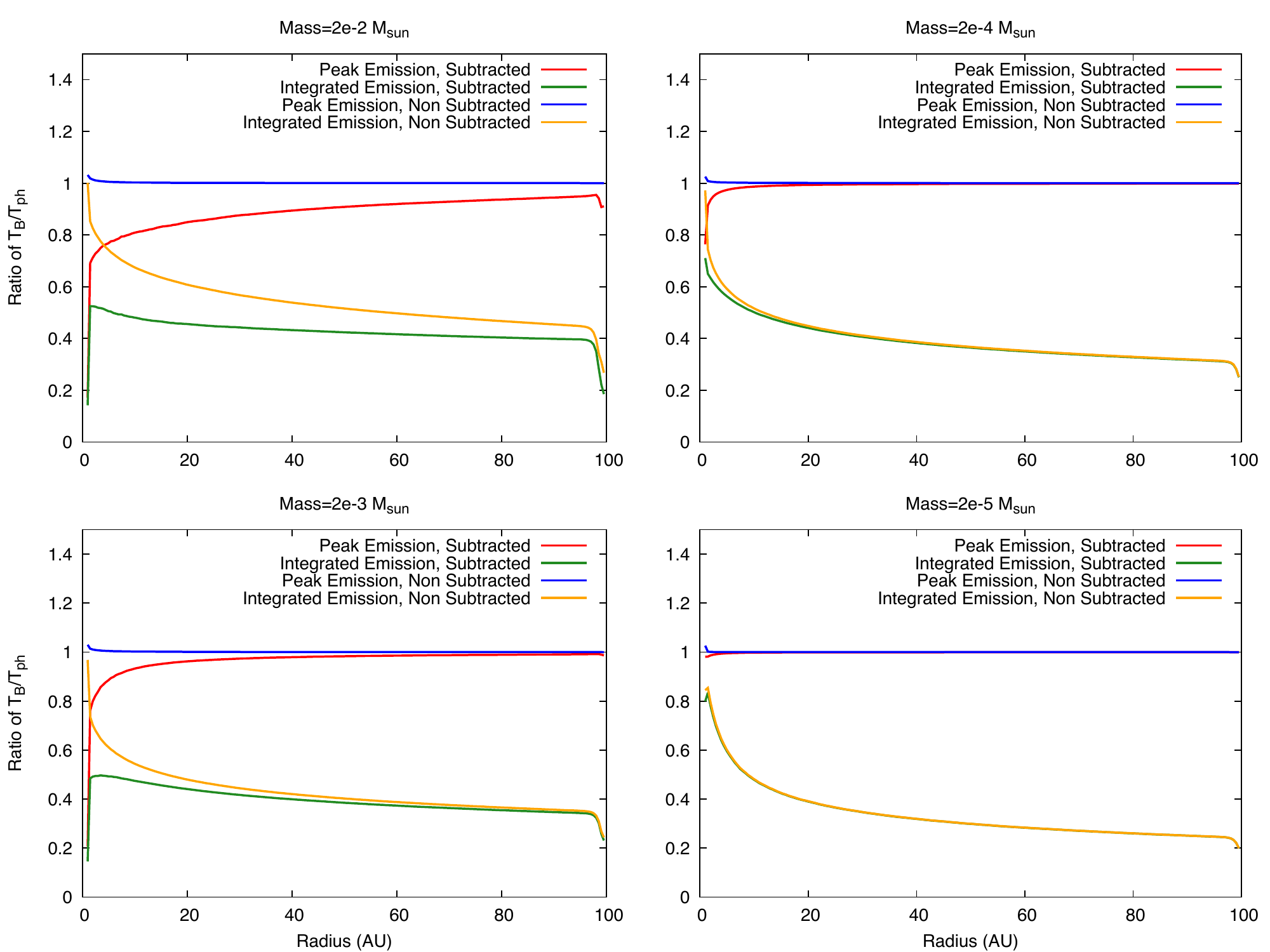}
\caption{\label{Fig:MoreDust} Relative accuracies of radial temperature for different masses of disk model for Integrated and 
Peak Emission Maps, for a variety of disk masses. For these models, the gas to dust ratio is much higher than the 
ISM value, set to 10 instead of 100. As expected, the effects of continuum subtraction are now more severe than 
before, especially for high mass models.}
\end{figure*}

Previous discussion has focused on models where the gas to dust ratio is 100, roughly the value of 
the nearby interstellar medium. However, it is likely that the ratio is not constant across all disks, or 
even across a given disk. (See for example \cite{Boe2017a, vdMa2016}, where the crescent visible 
in HD142527 and IRS 48 are found to contain far more dust than other regions of the disk.)

The effects of an altered gas to dust ratio are as follows: In cases where the gas to dust ratio is lower 
than the ISM value, the effects of continuum subtraction will be increased, while the effects of optically 
thin emission being included will not change. Lowering the gas to dust ratio increases the amount of 
dust present, which means that the background level of emission away from the line will increase. Thus, 
there is more emission which can be falsely attributed to continuum if the continuum is subtracted. 
However, until the continuum emission is high enough to be optically thick, the range of frequencies being 
integrated over will contain roughly the same fraction of optically thin and thick emission. 

Figure~\ref{Fig:MoreDust} shows the effect of assuming a gas-to-dust ratio of 10 (instead of 100) on the derivation 
of the disk temperature. While Figure~\ref{Fig:MoreDust} is qualitatively 
similar to Figure~\ref{Fig:Isothermal}, the effects of continuum subtraction are more exaggerated, particularly in the inner 
region of the disk. In the outer disk, deviation from the best measurable temperature is given primarily by 
inclusion of optically thin emission, and so doesn't show much change from the different gas to dust 
ratio.

The relation between disk physical temperature and line brightness temperature depends on the 
line optical depth, and therefore, from the abundance of the emitting molecules. $^{12}$CO is 
an abundant molecule in disks and its low-J transitions are optically thick even for relatively low CO 
column densities. CO isotopologues such as $^{13}$CO and C$^{18}$O are less abundant but still 
easily detected. If the disk is massive, even the line emission from CO isotopologues can be optically thick 
and allow to probe the disk temperature. C$^{18}$O, being more likely to be optically thin, can be used to 
determine gas masses. \citep[See][]{WiBe2014}.

\begin{figure*}[!t]
\centering{}
\includegraphics[width=\linewidth]{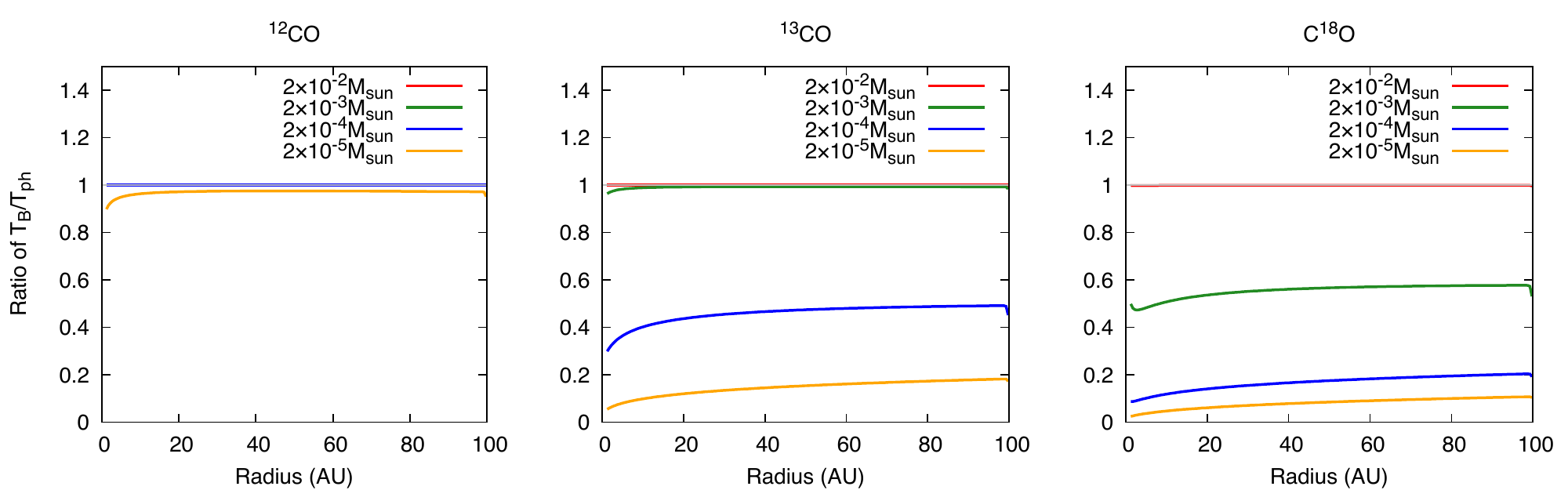}
\caption{\label{Fig:isot} Accuracy of measured brightness temperature for all four disk masses for the cases of $^{12}$CO, 
$^{13}$CO, and C$^{18}$O. In this case, the measured temperature is inferred via non-continuum subtracted peak emission map, 
since that technique has already been demonstrated to work best for disks whose emission is becoming optically thin.
}
\end{figure*}

Figure~\ref{Fig:isot} shows the ability to recover temperature for various disk masses for the cases of $^{12}$CO, $^{13}$CO, and 
C$^{18}$O. All temperatures are measured via non-continuum subtracted peak emission map, since that technique has already 
been demonstrated to work best for disks whose emission is becoming optically thin. Because $^{13}$CO and C$^{18}$O are much 
less abundant than $^{12}$CO, which limits their utility as temperature tracers, they offer consistently less useful results. $^{12}$CO, 
is optically thick at its peak for all disk 
masses covered here, except for the lightest most model, and even that case, the emission is still quite close to being optically thick. 
 $^{13}$CO is abundant enough to be useful for the heavier disk models, but by $2\times10^{-4} M_{\odot}$, the emission is too thin 
 to be useful. C$^{18}$O is uncommon enough that it can only be used for the very heaviest of disks.

An effect similar to that of an altered gas to dust ratio is the freezing out of CO molecules onto dust grains at low temperature. For CO, 
freeze-out typically takes place at temperatures of around 20K \citep[See][]{Ober2011}. This is a fairly low temperature, but is still reached 
by a fair fraction of the disk midplane. (See Figure~\ref{Fig:TempProf}, where the white line denotes the 20K contour.) The implications are 
as follows: For more abundant isotopologues, such as $^{12}$CO, there will be no noticeable effect, because the line becomes optically thick 
high enough in the disk atmosphere, that the freeze-out in the midplane is irrelevant. For less abundant isotopologues, such as C$^{18}$O, 
where the line becomes optically thick nearer the midplane (if it does at all), there will be a substantial loss of flux due to freeze-out. The 
primary emitting area, at the midplane, where density is highest, is depleted, and so the line is much less optically thick.

Because freeze-out simply removes molecules that would have emitted, it has the same effect as lowering the disk mass (or raising the 
gas to dust ratio). However, because of the temperature dependence, the effect is only significant for less common isotopologues. The 
primary hazard for observers regarding freeze-out is that a line which might be assumed to be optically thick based on the disk mass or 
brightness of the $^{12}$CO line, may not be because all of the CO at the midplane is frozen out.

There are additional subtle effects which can also alter the abundances of CO and its isotopologues. Possible depletion can come from 
active disk chemistry.  \citep[See][]{Kama2016}. Another such effect is CO self shielding.  \citep[See][]{Schw2016}.

\section{Physical Disk Model}

\subsection{Structure and Emission}

\begin{figure}[!t]
\centering
\includegraphics[width=\linewidth]{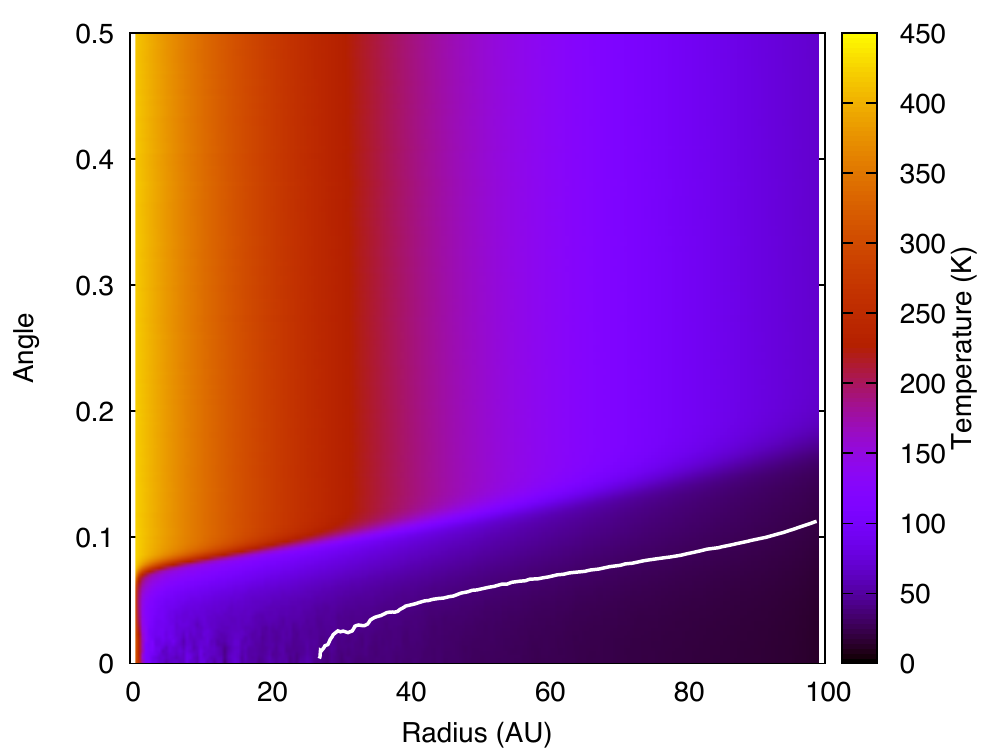}
\caption{\label{Fig:TempProf} Example temperature profile for a disk of mass $2\times10^{-2}M_{\odot}$, as calculated using RADMC3D. 
The angle is measured relative to the midplane. The radial temperature profile is similar to previous models, but now 
the vertical temperature gradient varies, starting at the very cold midplanar regions, and heating up at the lower density, 
high altitudes. The temperature profile is symmetric about the midplane. Separate temperature models were generated 
for each disk mass, however, they are qualitatively highly similar. The white line indicates the 20K contour. Inside this 
region, CO freezes out onto dust grains.
}
\end{figure}

The models discussed above contain all of the relevant features of
protoplanetary disks, but are nonetheless extremely simple models,
particularly regarding the vertical temperature profile. A more realistic temperature
profile for the outer regions of an externally illuminated disk would have a colder midplane 
and hotter disk surface, assuming that viscous heating is negligible. \citep[See][]{Dull2002}. Such a temperature structure can be obtained 
by solving the radiative transfer equations taking into account all 
the heating and cooling sources, as well the vertical disk structure. Here, we use the  
Monte Carlo radiative transfer code RADMC3D \citep{Dull2012}, to calculate the temperature distributions $T(r,z)$ 
of disks characterized by the same surface densities and masses discussed in the previous section. Figure~\ref{Fig:TempProf} 
shows an example temperature profile for a disk of mass $2\times10^{-2}M_{\odot}$. However, it is worth mentioning that high in 
the disk atmosphere, the temperature of gas and dust may no longer match. When $n < n_{crit}$ (see equation 6), the gas is no longer in 
LTE, and can be much hotter than the dust. \citep[See][]{Berg2016}.

Because of the vertical temperature profile, it is no longer possible to determine the dust and gas intensity 
analytically, and it is also no longer possible to simply describe the ratio between line brightness temperature and 
physical disk temperature as done in the previous session. However, we can use ray tracing calculations 
to investigate the accuracy of the inferred gas temperature from the line emission. To this end, we created 
our own ray tracing algorithm which allows us to numerically calculate the pure line and continuum emission. 
This algorithm differs from the ray tracing module of RADMC3D in that it numerically integrates the line 
and continuum radiative transfer equation (Equation 5) without the need to specify a three dimensional grid, and 
it is written to use parallelization to speed up the ray tracing. The details of the
ray tracing code and a comparison with RADMC3D are presented in Appendix C.

\begin{figure*}[!t]
\centering{}
\includegraphics[width=\linewidth]{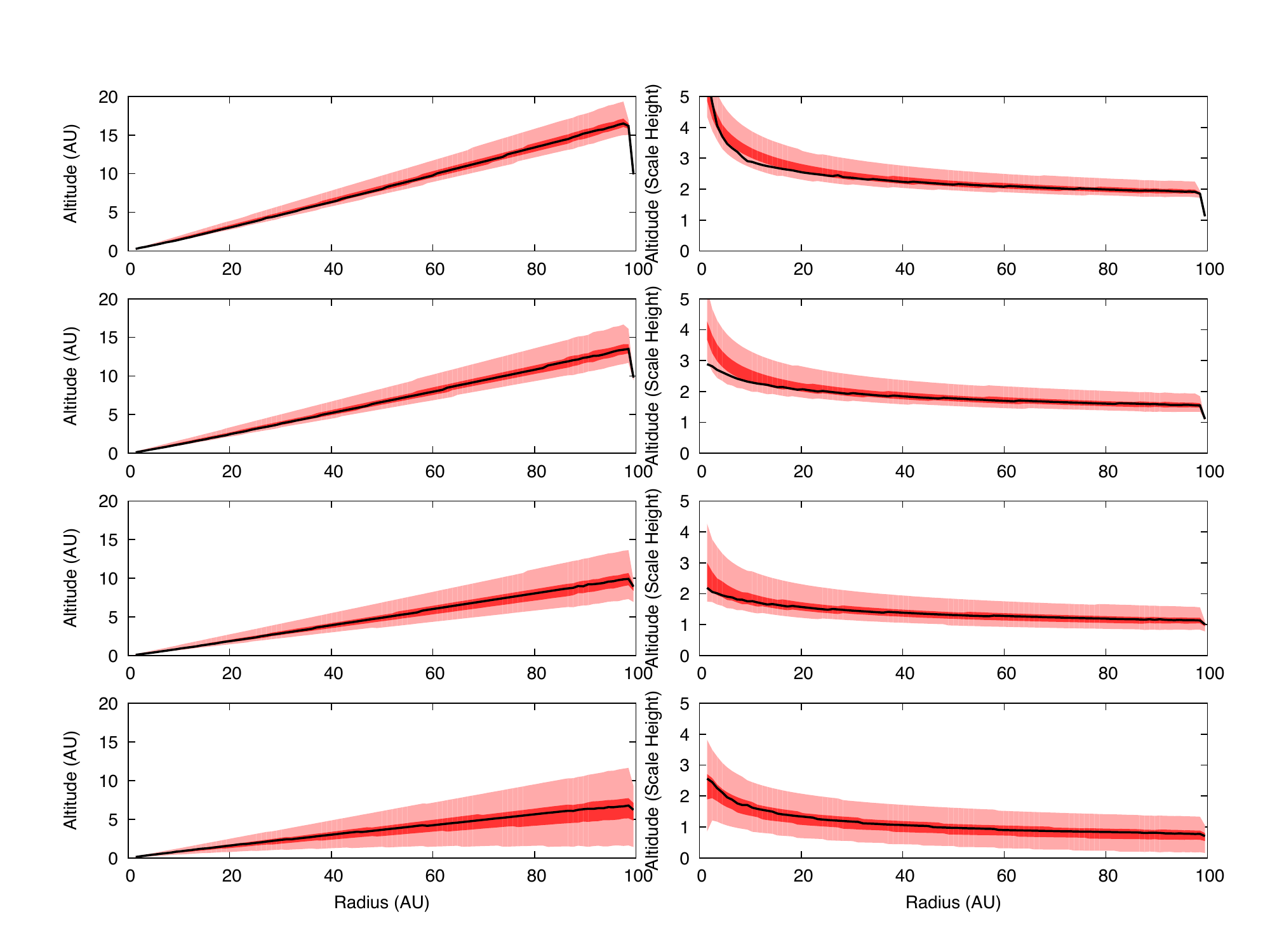}
\caption{\label{Fig:EmReg}Regions dominating disk emission for face-on disk models with masses from 
$2\times10^{-2}M_{\odot}$ to $2\times10^{-5}M_{\odot}$. The left column shows the emission region in the 
RZ plane. The right column displays the same data, but with the altitude in terms of scale heights, not 
astronomical units. The lightly shaded region corresponds to the region producing 90\% of the emission, 
while the darker shaded region corresponds to the inner 33\% of the emission. The line represents the 
height at which the observed brightness temperature matches the physical temperature of the disk.
}
\end{figure*}

In the preceding sections, we established that the non continuum subtracted peak emission of an optically thick line 
is the best probe of the temperature of the emitting gas. However, if the disk is not vertically isothermal and/or not 
observed at face-on inclination, the brightness temperature of the line would correspond to an average temperature 
over the region where the line becomes optically thick, rather than to a specific point. It is therefore important to consider 
the morphology of the emitting region. Figure~\ref{Fig:EmReg} shows the location of the disk regions emitting the $^{12}$CO J=3-2 as
observed at face-on inclination. The lightly shaded region of each panel represents the region responsible for 90\% of 
the line emission, while the more heavily shaded area corresponds to the region producing the inner 33\% of the 
line emission. The line represents the height at which the observed brightness temperature matches the physical 
temperature of the disk.

The position where brightness temperature matches the physical temperature of the gas for $^{12}$CO 
depends on the optical depth of the line, and therefore on the disk mass. For disks more massive than $M=2\times 10^{-3}$ M$_\odot$, 
the point where the brightness temperature is equal to the physical temperature of the gas is located between two and three scale heights. 
At lower masses, the location moves towards the midplane.
The height of the emitting layer increases from about 1$h$ for $M=2\times 10^{-5}$ M$_\odot$
to 3$h$ for $M=2\times 10^{-2}$ M$_\odot$. Consequently the temperature of the line will decrease with the disk mass from a value similar 
to that of the disk surface, which is typically located between four and five scale heights \citep[See e.g.][]{Dull2001}, to that of the disk midplane. 

The point of matching temperature is roughly in the middle of the emission region, as expected. The width of the emission region has important implications for determining 
vertical temperature gradients for protoplanetary disks. At around a scale height, the emission region is thick, and represents a 
fair fraction of the disk area. Because of this, lines of lower optical depth can overlap, complicating the derivation of the 
vertical temperature structure.

\subsection{Derivation of Gas Temperature from the Line Emission}

Following the analysis presented in the previous session, we first decompose the line emission in true gas
emission, i.e., line emission attenuated by dust absorption, and true dust emission, i.e., continuum emission attenuated
by gas absorption. Due to the vertical temperature gradients, Equation 11 no longer holds, and these two terms must 
be calculated numerically. We do that by performing two ray tracing simulations: The first includes both gas and dust opacities 
to calculate the combined emission. In the second, we only use the dust opacity in the source function term, and both dust and 
gas opacities in the attenuation term of the radiative transfer equation. These two emissions are then subtracted to obtain 
the true gas emission, which is finally converted in a true brightness temperature using Equation 15. 
If the line is optically thick, the true line brightness temperature will correspond to a weighted average of 
the physical gas temperature across the region emitting the line (see Figure~\ref{Fig:EmReg}). 
Instead, if the line is optically thin, the true line brightness temperature will be lower than the temperature of the 
emitting layer.

\begin{figure*}[t]
\centering
\includegraphics[width=\linewidth]{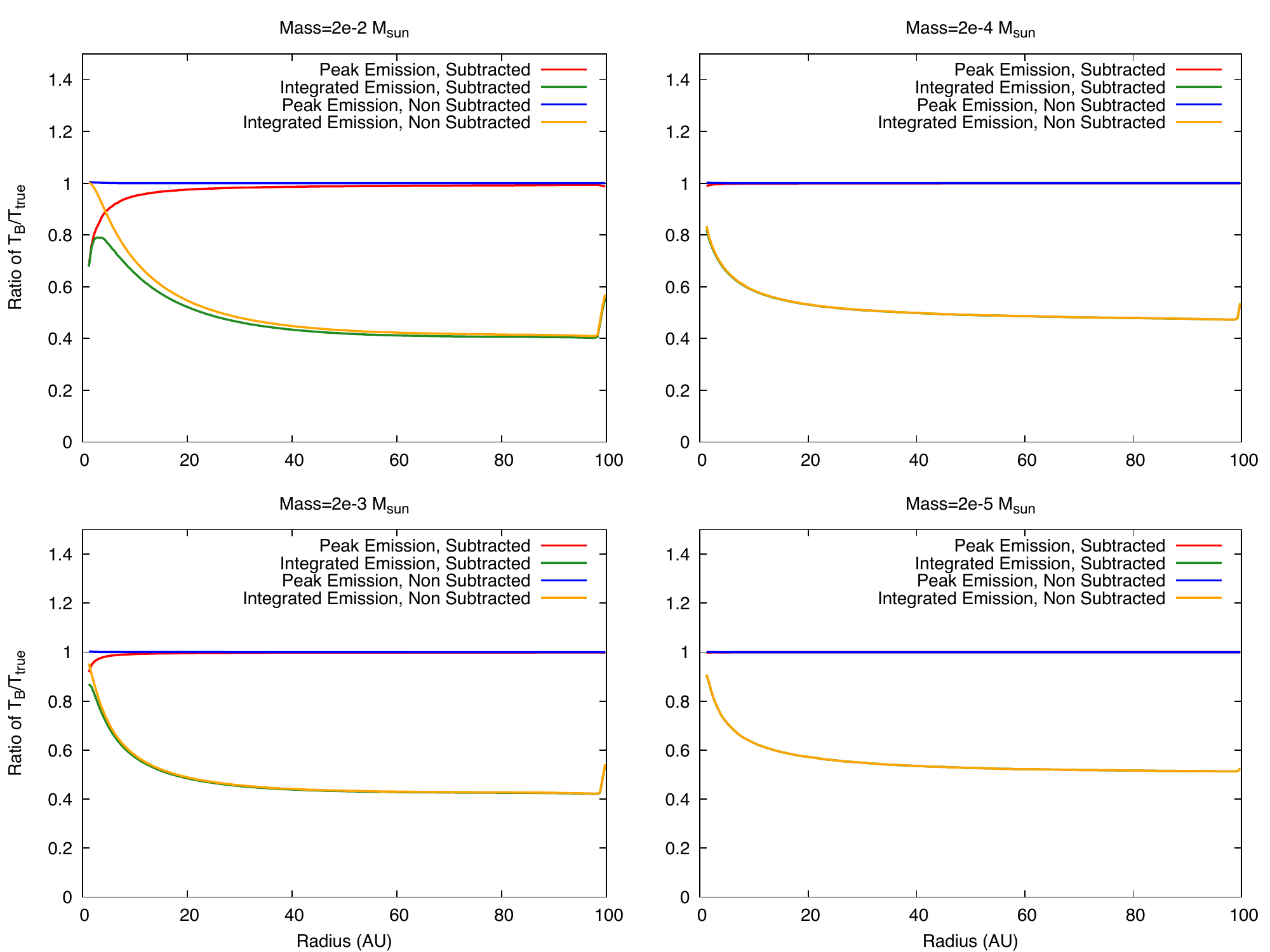}
\caption{\label{Fig:NonIsothermal} Relative accuracies of radial temperature for different masses of disk model for Integrated Emission Map, and for Peak Emission Map. In these models, the disk is no longer vertically isothermal. Since there is no longer a single physical temperature to recover at each radius, the comparison is against the most accurate possible brightness temperature which can be recovered.}
\end{figure*}

As before, we use the ratio between the brightness temperature of the line derived from both peak and  
integrated emission, both with and without continuum subtraction, and the true brightness temperature, 
to investigate the most reliable method for deriving gas temperatures. 
Figure~\ref{Fig:NonIsothermal} shows this ratio as a function of radius for the same disk models shown in Figure~\ref{Fig:Isothermal}.
The difference is that the disks are no longer vertically isothermal. 
However, as in the vertically isothermal case, the non continuum subtracted peak emission map is the best probe of 
gas temperatures, while continuum subtraction leads to underestimating the gas temperature in the 
regions where the continuum is optically thick. This effect is less severe than in the isothermal case, because
the disk midplane, which is where most of the dust emission originates, is colder than the layer emitting the line. 
The brightness temperatures derived from the continuum subtracted integrated intensity (yellow lines) is, 
on average, half of the gas temperature and should therefore not be used for this purpose. 

It is worth noting one crucial difference between vertically isothermal and non isothermal models. The temperature derived from the 
non continuum subtracted peak intensity in the non-isothermal model appears to trace the gas 
temperature much better than in the isothermal model, particularly at low mass and high radius. This is a result of the fact that 
while in Figure~\ref{Fig:Isothermal} we compare the line brightness temperature to the real gas temperatures, 
in Figure~\ref{Fig:NonIsothermal} we no longer can, because the model is not vertically isothermal, and 
no single temperature characterizes a given radius. Instead, we compare the line brightness temperature with the 
true brightness temperature calculated by subtracting the attenuated continuum model from the full model. 
In the case of optically thick emission, the true brightness temperature is equal the physical temperature of the emitting 
region, but for the lower mass, and more optically thin models, both the true brightness temperature and the brightness temperature 
inferred from the observations will be lower than the gas temperature. 

\begin{figure*}[!t]
\centering
\includegraphics[scale=0.5]{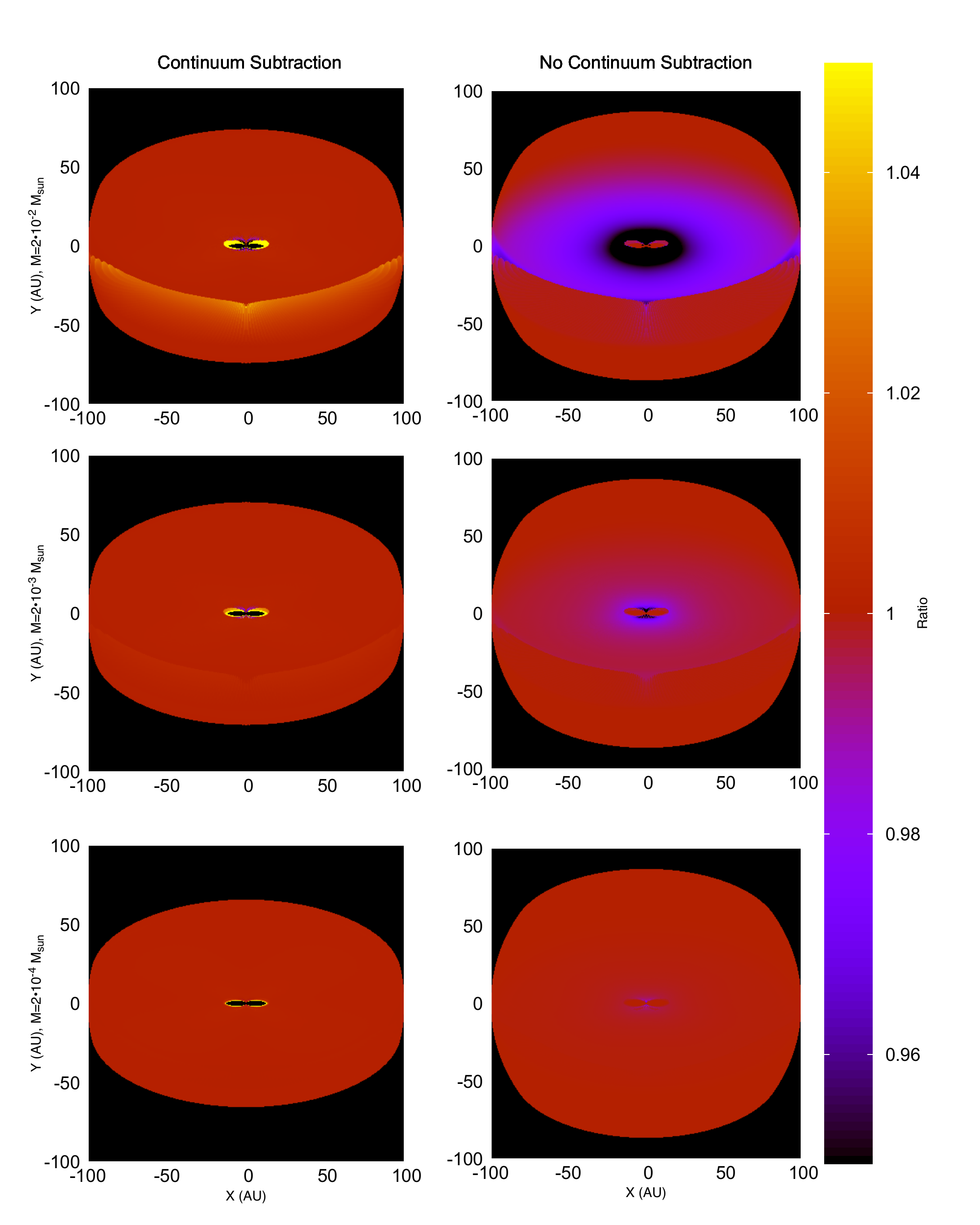}
\caption{\label{Fig:Inclined}Accuracy of brightness temperature measured via Peak Emission Map for masses 
ranging from $2\times10^{-2}M_{\odot}$ to $2\times10^{-4}M_{\odot}$, with and without continuum subtraction,
for an inclination of 60°. Deviation from the best recoverable brightness temperature 
is greatest towards the center, and along the surface where the tilted disk most increases the line 
of sight. The deviation is only of order a few percent.
}
\end{figure*}

In the disk models presented here, the optical depth of the $^{12}$CO 3-2 line emission at the center of the line is 
about $3.6\times10^4$, at the inner radius of the most massive disk, is about 10 at 60 AU in the $2\times10^{-5}$ M$_\odot$ disk 
model, and never quite reaches unity, even for the lightest model. This comes primarily from $^{12}$CO being a very 
abundant molecule, and the 3-2 line being quite bright at these temperatures. This means that the line emission remains 
optically thick ($\tau > 1$) for the peak of the line even at CO column densities of $4.2\times10^{-7}$ g cm$^{-2}$, 
or number column densities less then $1.8\times10^{16}$  cm$^{-2}$. The location at which the line brightness temperature inferred from the 
non-continuum subtracted peak emission is equal to the gas temperature is indicated by black curves in Figure~\ref{Fig:EmReg}. When the line is optically thick, the brightness temperature of the line probes the temperature 
of a disk layer included in the region that emits most of the line. As the line becomes optically thin, the brightness 
temperature decreases until it drops below the temperature of the gas on the disk midplane, which is the lowest gas 
temperature value in each vertical column. In the face on case, measurements of optically thick emission lines characterized 
by different optical depths could therefore allow to estimate the three dimensional disk vertical structure if the optical depth 
of the line is known. 

\subsection{Effect of Inclination}

For ease of analysis, the preceding discussion has considered only disks which 
are viewed face on. However, most actual observations must deal with some degree
of inclination. The primary effects of tilting the disk are as follows: First, Keplerian 
rotation will introduce a velocity component along the line of sight, with the result that 
for a given frequency window, only a small region will be emitting strongly. Because of 
this, the morphology of the spectrum changes significantly. Instead of being a 
relatively simple, singly peaked, distribution, the spectrum becomes much wider 
and might show two peaks symmetric with respect to the center of the line. 
In terms of measuring a brightness temperature, however, there is little change from the 
face-on case. The spectrum for an inclined disk is more spread out in frequency 
than the sharply peaked spectra from face-on disks, and has larger contributions from 
optically thin regions. This implies that temperatures derived from the spectrally 
integrated measurement will underestimate the gas temperature by an even larger amount 
compared to the face-on case. Therefore, in order to more concisely discuss the effect of 
the disk inclination on the determination of the gas temperature, in this section we will restrict 
the analysis to the line peak emission.

\begin{figure*}[!t]
\centering
\includegraphics[width=0.48\linewidth]{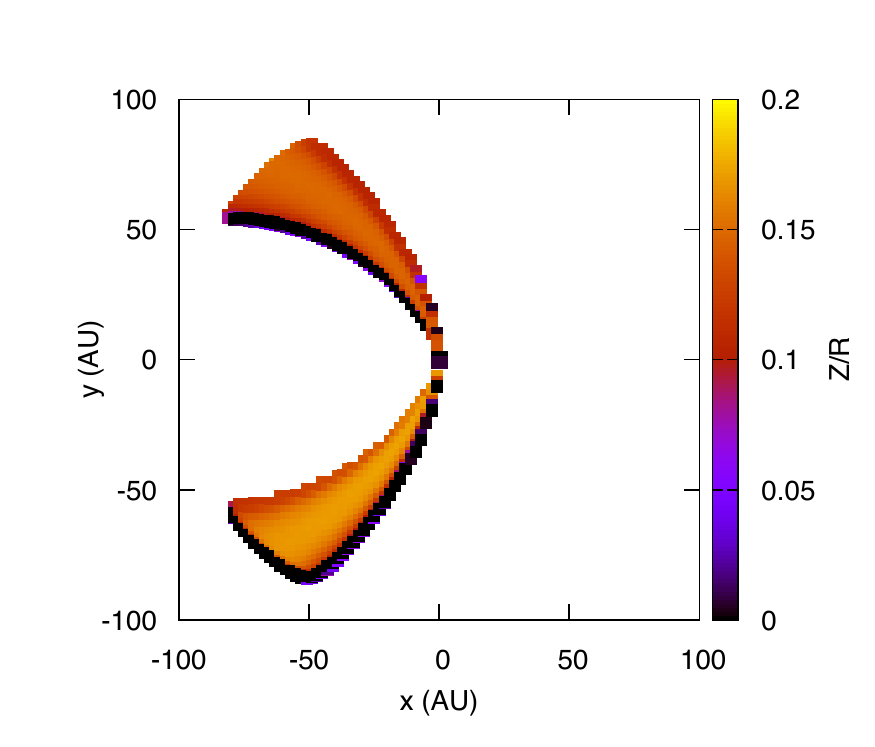}
\includegraphics[width=0.48\linewidth]{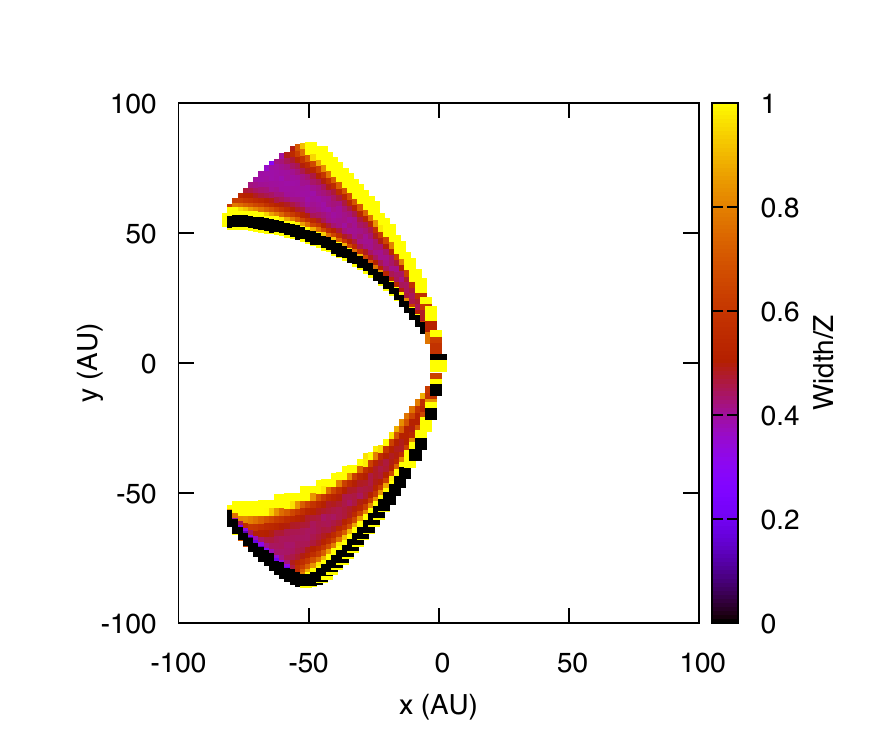}
\caption{\label{Fig:EmRegInc} The emission region of a disk inclined at an angle of 60°, for a mass of $2\times10^{-3} M_{\odot}$. 
The emission comes only from the $^{12}$CO 3-2 line, measured 1 km sec$^{-1}$ off of line center. Left: Each point represents 
the height of the center of the emission region for that point in the disk, shown as Z/R. Right: The width of the emission 
region, divided by the height of the center at each point. At points of high optical depth, the center of the emission region is 
higher above the disk midplane, and the region is narrower.}
\end{figure*}

Because of the non-face on inclination, the disk emission is no longer azimuthally symmetric and 
we can no longer plot temperatures as a function of the orbital radius. 
Instead, we show in Figure~\ref{Fig:Inclined} two-dimensional images of 
 the ratio between the true brightness temperature and the line brightness temperature derived 
 from peak intensity maps. The models correspond to disk masses of $2\times10^{-2}$ M$_\odot$, $2\times10^{-3}$ M$_\odot$, 
 $2\times10^{-4}$ M$_\odot$, observed at an inclination of 60\arcdeg. 

Without continuum subtraction, the line brightness temperature is slightly higher than the true 
brightness temperature closer to the center of the disk and along the disk outer edge. 
This is because tilting the disk increases the length of the lines of sight through 
the disk, causing optical depth of both gas and continuum emission to increase. 
This effect increases as inclination increases, and it is most noticeable in the heaviest model 
with mass $2\times10^{-2}$ M$_\odot$, as the effect is most 
pronounced where dust emission is more significant. Lighter models are largely 
unaffected, because the disk is so light that even in the denser inner regions, the dust emission is so negligible that there simply 
isn't anything to falsely attribute to the line. In any case, including the continuum emission in estimating 
the line temperature leads to errors smaller than 5\%.  

If the continuum is subtracted, as shown in the left column of Figure~\ref{Fig:Inclined},  
the error goes in the opposite direction, and emission is underestimated. The underestimate 
from continuum subtraction is larger than the overestimate obtained from not subtracting continuum 
and can be as high as 10\%-15\% in the most massive case. 
Recall, however, that the apparent accuracy in this case 
is only relative to the true brightness temperature, and that for optically thin emission,
the brightness temperature is far below the gas temperature. 

This example considered a disk inclined by 60°. Because the disks are geometrically thin, the lines of sight will 
lengthen as the disk is inclined. This means that the effects described above will be magnified for higher 
inclinations.

Finally, it is important to consider the emission region of the inclined disk, as was done for the face-on models. In the 
more geometrically complicated inclined case, the lines of sight are no longer straight down, and the results are not 
radially symmetric. Additionally, the Keplarian rotation of the disk means that only a fraction of the gas is emitting at the 
right frequency.

Figure~\ref{Fig:EmRegInc} shows two maps of the emission region for the inclined disk of mass $2\times10^{-3} M_{\odot}$,
for the $^{12}$CO 3-2 line, imaged at one km $s^{-1}$ off of line center. The left panel shows the height of the emission region (defined
as the height above the midplane of the centroid of each line of sight), divided by the radius. The right panel 
shows the width of the emission region, defined as the region producing the middle 90\% of the emission, divided by the height 
of the center. Because of the Keplarian rotation of the gas, most of 
the gas picks up a line of sight velocity, and the line emission is doppler shifted out of the frequency window. Only a small region 
still emits at the correct frequency. This region is dominated by CO emission, which is optically thick, and 
the emission region behaves similarly to in Figure~\ref{Fig:EmReg}. The center of the emission region is located above the midplane, 
because the CO emission quickly becomes optically thick. The width of the emission region is small roughly 30-40\% of the width 
away from the center, because the optical depth rises quickly. As the line weakens, away from the y axis, the width increases 
as the optical depth drops. The width also rises further away from the center, for the same reason. As disk mass increases, the emission 
region shifts to higher altitude and becomes narrower.

\section{Derivations of the Gas Temperature from Synthetic Observations of a Physical Disk Model}

Lastly, it is key to consider the effects that a real observation would have on recovering an accurate 
line brightness temperature and, ultimately, the gas temperature. The discussion thus far has focused 
on synthetic disk models, and has neglected that interferometric observations are affected by finite noise and resolution, 
as well as spatial filtering. In this section, we first investigate the effect of the finite angular resolution by convolving synthetic disk models of the 
CO emission with 2-dimensional Gaussian functions that mimic  ALMA point spread functions. We then investigate the combined
effect of noise, angular resolution, and spatial filtering by producing synthetic ALMA observations of our disk 
models. 

\subsection{Effect of Beam Convolution}

\begin{figure}[!t]
\centering
\includegraphics[width=0.9\linewidth]{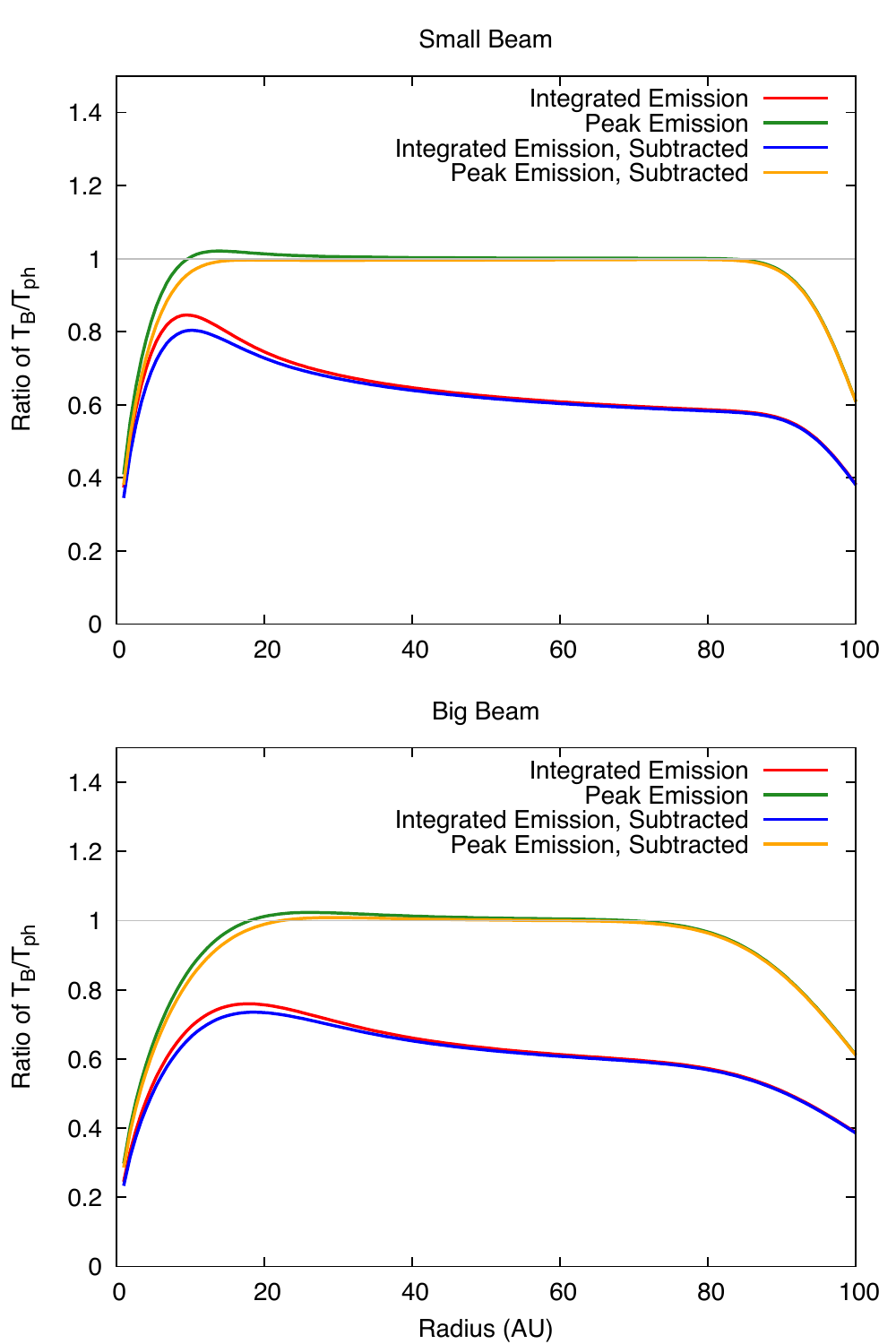}
\caption{\label{Fig:BlurRad} Ratios of measured brightness temperature to physical temperature for a vertically 
isothermal, face on disk with a mass of $2\times10^{3}M_{\odot}$. Top panel: models convolved with a 0.11\arcsec$\times$0.09\arcsec\ beam. 
Bottom panel: models convolved with a 0.16\arcsec$\times$0.14\arcsec\ beam. In both cases, beam dilution primarily causes problems in the 
inner and outermost regions of the disk.}
\end{figure}

We first investigate the effect of beam convolution on the emission from the pedagogical disk model discussed in 
Section~2. Though not realistic, this model has the advantage of allowing us a precise comparison between 
line brightness temperature $T_B$ and gas temperature $T_{ph}$. Figure~\ref{Fig:BlurRad} shows the radial profile of the 
ratio $T_B/T_{ph}$ corresponding to the CO emission from the $2\times10^{-3} M_{\odot}$ disk model convolved with 
Gaussian beams of 0.16\arcsec$\times$0.14\arcsec\ and 0.11\arcsec$\times$0.09\arcsec\ (top and bottom panel, respectively).  
This case is analogous to the lower left panel of Figure~\ref{Fig:Isothermal}. 
Convolution with the beam smears out emission, and has the most effect where the intensity 
changes rapidly. Thus, the inner and outer edge of the disk are strongly 
affected, and the ability to determine the correct brightness temperature is reduced. Beam dilution is the strongest within 
half a beam from the center, and therefore it becomes more severe as the beam size gets larger. 
The problem of beam dilution is independent of the choice of peak emission map or integrated emission map 
and the choice of continuum subtraction. All methods are affected equally.  

\begin{figure*}[!t]
\centering{}
\includegraphics[width=\linewidth]{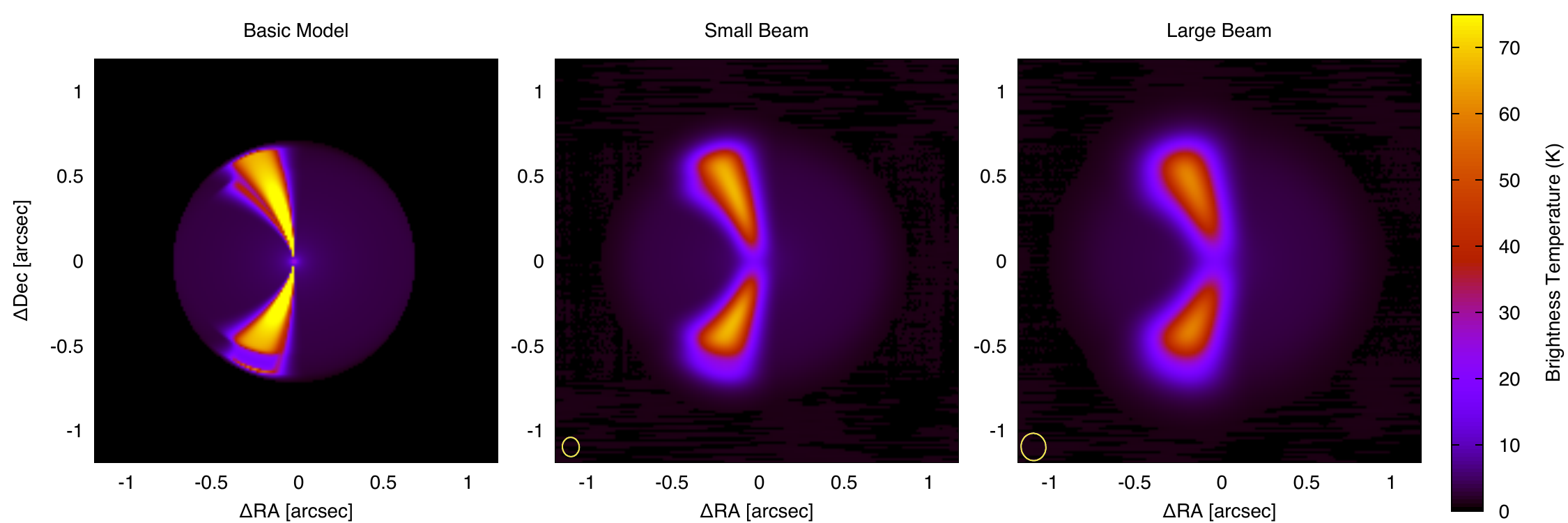}
\caption{\label{Fig:BlurChannel} The effect of the beam on a single channel map. From left to right: A single channel map, not 
convolved with a beam, the same channel map, but now convolved with a small beam, and lastly, the channel map 
convolved with a large beam. The beam is indicated by the ellipse in the lower left corners of the middle and right 
panels. Color indicates brightness temperature in Kelvin.}
\end{figure*}

To further demonstrate the effects of beam convolution, we preset in Figure~\ref{Fig:BlurChannel} 
synthetic brightness temperature maps of the $^{12}$CO J=3-2 emission from the 
$2 \times 10^{-3}$ M$_\odot$ non vertically isothermal disk model observed observed at an 
inclination of 30\arcdeg. The maps correspond to the line plus continuum emission calculated at  
a velocity of 1 km s$^{-1}$ and show the effect of the convolution with beams of 0.11\arcsec$\times$0.09\arcsec\ and 
0.16\arcsec$\times$0.14\arcsec\ in size. 

Due to the effect of beam dilution, the brightness temperature of the line 
decreases with increasing beam sizes. This effect is the strongest where the emitting area is the 
smallest, as in the central region of the disk where, due to Keplerian rotation, the emitting region in 
a given channel becomes extremely narrow. This means that very small beam sizes are required to measure gas temperatures 
in the innermost disk regions. In the case of a beam size of the 0.11\arcsec$\times$0.09\arcsec\ beam, the brightness 
temperature of the line reaches a maximum of 67.3 K, while the true brightness temperature in the same pixel is 
73.7 K. However, in the more strongly affected central disk region of the channel map, the true brightness temperature reaches a 
maximum of 159.1 K, but the beam-convolved model only has a value of 23 K. This implies that beam dilution has a strong 
effect in measuring gas temperatures in protoplanetary disks, even 
at the resolutions provided by ALMA. Clearly, ALMA is capable of even higher resolution. However, since the integration 
time required to achieve a given signal-to-noise ratio increases as $\theta^{-4}$, observations of the line emission at a 
resolution smaller than about 0.1\arcsec\ requires a considerably large amount of telescope time. 

Beam dilution affects integrated emission map and peak emission map differently. 
In particular, our simulations show that the spectrally integrated emission map is strongly 
resilient against the effects of beam dilution. The reason for this is that although the beam convolution smears 
out the emission, the end result is that flux lost from one resolution element is moved in the nearby resolution element. 
As a consequence, beam dilution leads to a spectral broadening of the line such that despite the fact that the peak 
emission might substantially drop, the integral is roughly the same. 

\begin{figure}[!t]
\centering
\includegraphics[width=\linewidth]{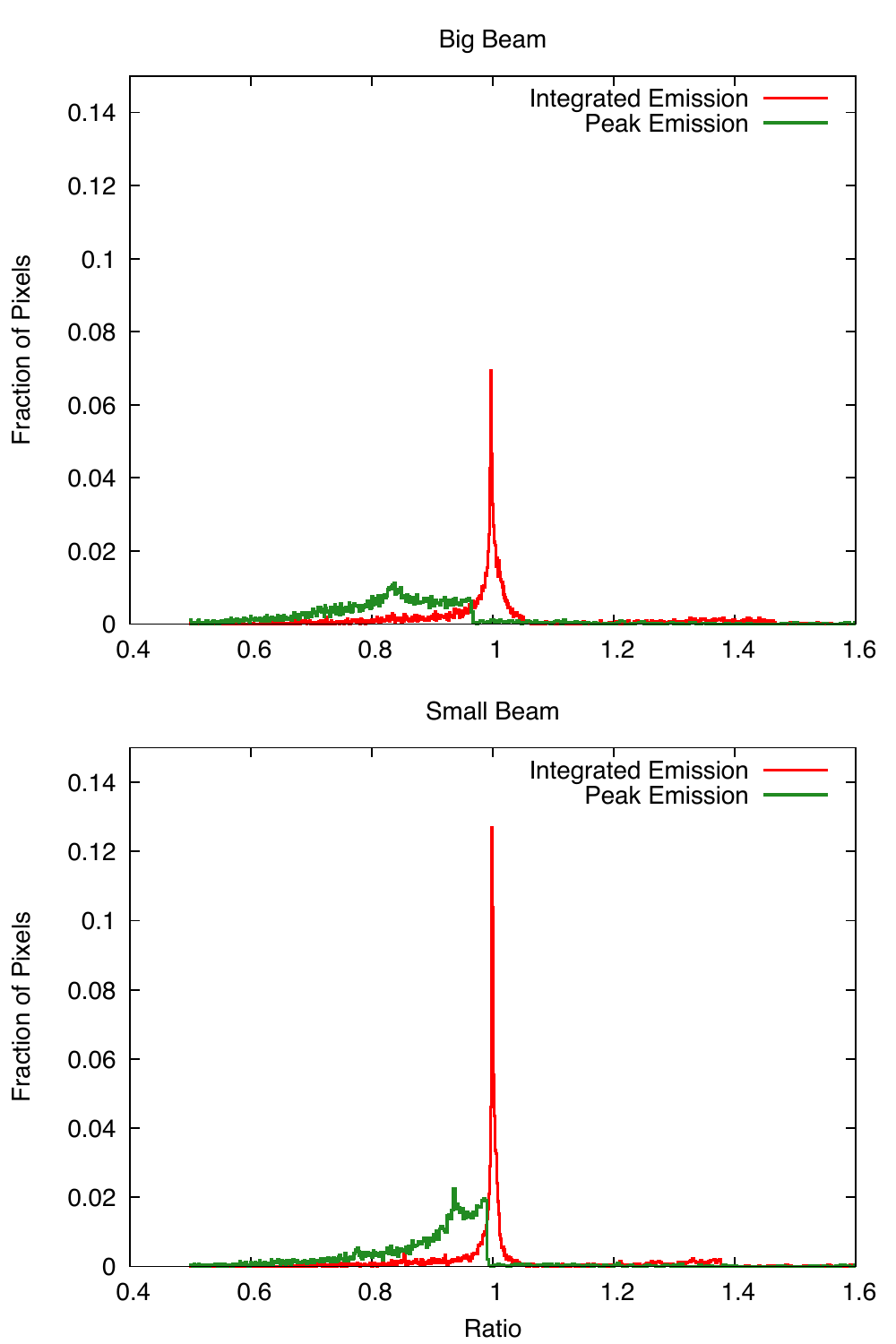}
\caption{\label{Fig:hist1} Top Panel: Ratio of beam convolved model to non-convolved model for the large beam size. A ratio of 1 
indicates that the convolution has no effect, while a higher ratio means that the convolved model has lost flux. While 
the integrated emission map is largely unaffected, the peak emission map is hardly usable. Beam dilution has removed 
much of the flux from the image, yielding a brightness temperature artificially low by, on average, 20\%. Bottom Panel: Ratio 
of beam convolved model to non-convolved model for the smaller beam size. Like before, the 
integrated emission map is mostly unaffected. Now, however, with a smaller beam, the effects on the peak emission map 
are better controlled, though still not as good as the integrated emission map. The average loss of flux is now only on the 
order of 5-10\%.}
\end{figure}

\begin{figure}[!t]
\centering
\includegraphics[width=\linewidth]{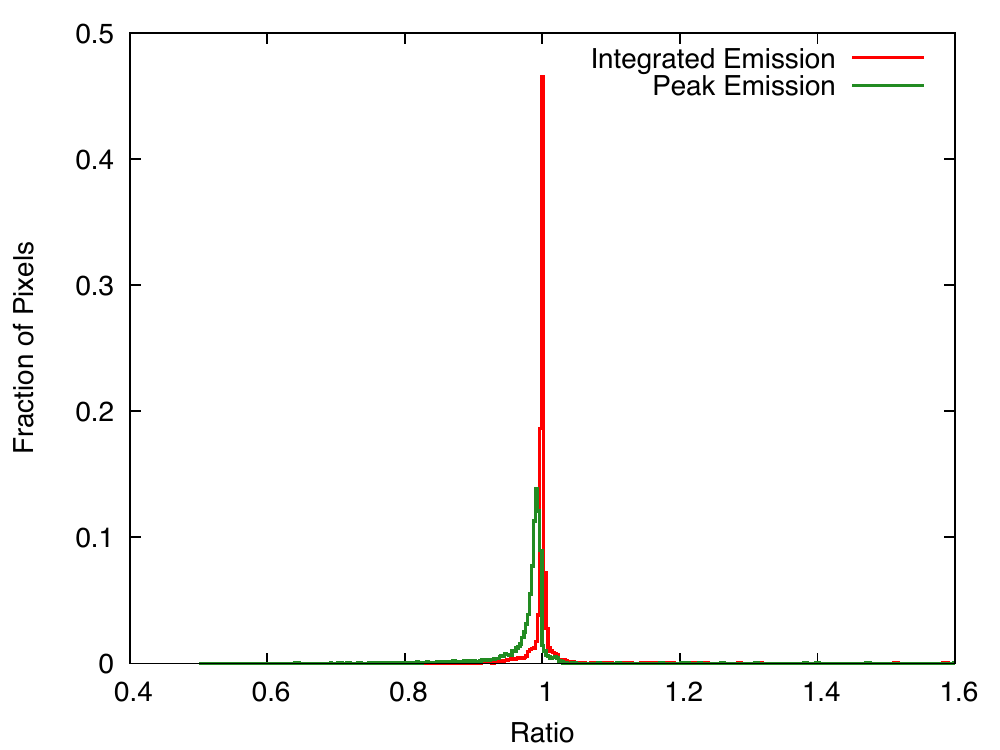}
\caption{\label{Fig:hist2}Ratio of beam convolved model to non-convolved model for the smaller beam size, for a larger disk 
with radius of 300AU. In this case, the beam size is small enough compared to the channel map scale, that both methods 
adequately recover the non-convolved temperature.}
\end{figure}

This conclusion is supported by Figure~\ref{Fig:hist1} which shows histograms  of the ratio 
between the line brightness temperature derived both from integrated emission and peak emission maps, and the
true line brightness temperature, for the disk model discussed above convolved with beam sizes of 0.11\arcsec$\times$0.09\arcsec\ 
and 0.16\arcsec$\times$0.14\arcsec. In both cases, the curves corresponding to the temperature derived from integrated 
emission maps show a strong peak around 1, with 90\% of the values having ratios between 0.73 and 1.31 for the larger beam, 
and 0.77 and 1.27 for the smaller beam.
Temperature derived from the peak emission map are instead spread across a large range of values. 
This is because most of the inner region of the disk has lost flux because of beam dilution, and the recovered 
brightness temperature is low by an average of 10\%-20\%, but with some pixels being much worse. This demonstrates the 
primary constraint on the use of the peak emission map method: the beam size must be small compared to the size of the 
channel maps in order for the results to be worthwhile. 
The disk considered in Figure~\ref{Fig:hist1} is only 100 AU in radius, making it fairly small. Figure~\ref{Fig:hist2} instead 
shows the effects of observing a larger disk model, where the radius has been increased to 300AU, and the smaller beam size. 
This means that the channel map size is now much larger than the beam, and both methods work almost equally well.

\subsection{Effect of finite sensitivity}

The other issue introduced by actual observation is that of noise. Spatially, noise is typically on the scale of the beam size, 
and its magnitude depends primarily on the length of the observation. In order to simulate the noise of a real observation 
such as those described in the previous section, models are processed using the {\it simobserve} function of CASA. 
For sake of simplicity, we simulated ALMA observations of only the disk model with a mass of $2\times10^{-3}$ M$_\odot$ (that of 
Figure~\ref{Fig:Inclined}, \ref{Fig:EmRegInc}, and \ref{Fig:BlurChannel}). 
As before, we placed the disk a distance of 140 pc from Earth at a declination of -24\arcdeg, and we inclined it by 30\arcdeg\ 
with respect to the line of sight. Using our
 ray tracing code, we generated channel maps of the $^{12}$CO 3-2 plus dust emission emission between $\pm$21.69 km s$^{-1}$ with a 
 velocity resolution of 0.0145 km s$^{-1}$. The synthetic channel maps were then observed with the {\it simobserve} task of CASA 
 adopting the antenna configurations 14 and 17, which deliver beam sizes of 0.16\arcsec$\times$0.14\arcsec\ (for 1hr synthesis 
 and natural weighting), and 0.11\arcsec$\times$0.09\arcsec\ (for 5hr synthesis and natural weighting), respectively.  
 The largest recoverable angle scale at the frequency of the line for the chosen configurations is 7.28\arcsec and 3.63\arcsec, respectively, 
 indicating that the size scale of the disk is smaller than the largest recoverable structure, and no flux is lost. 
 Thermal noise typical of ALMA weather conditions was included resulting in an rms noise value per channel 
of 4.2 mJy beam$^{-1}$ in the 1~hr long track, and 1.9 mJy beam$^{-1}$ in the 5~hr long track 
CO emission is detected with a peak signal-to-noise ratio of 24 in the higher resolution image and 19 in the lower resolution image. 

\begin{figure}[!t]
\centering
\includegraphics[width=\linewidth]{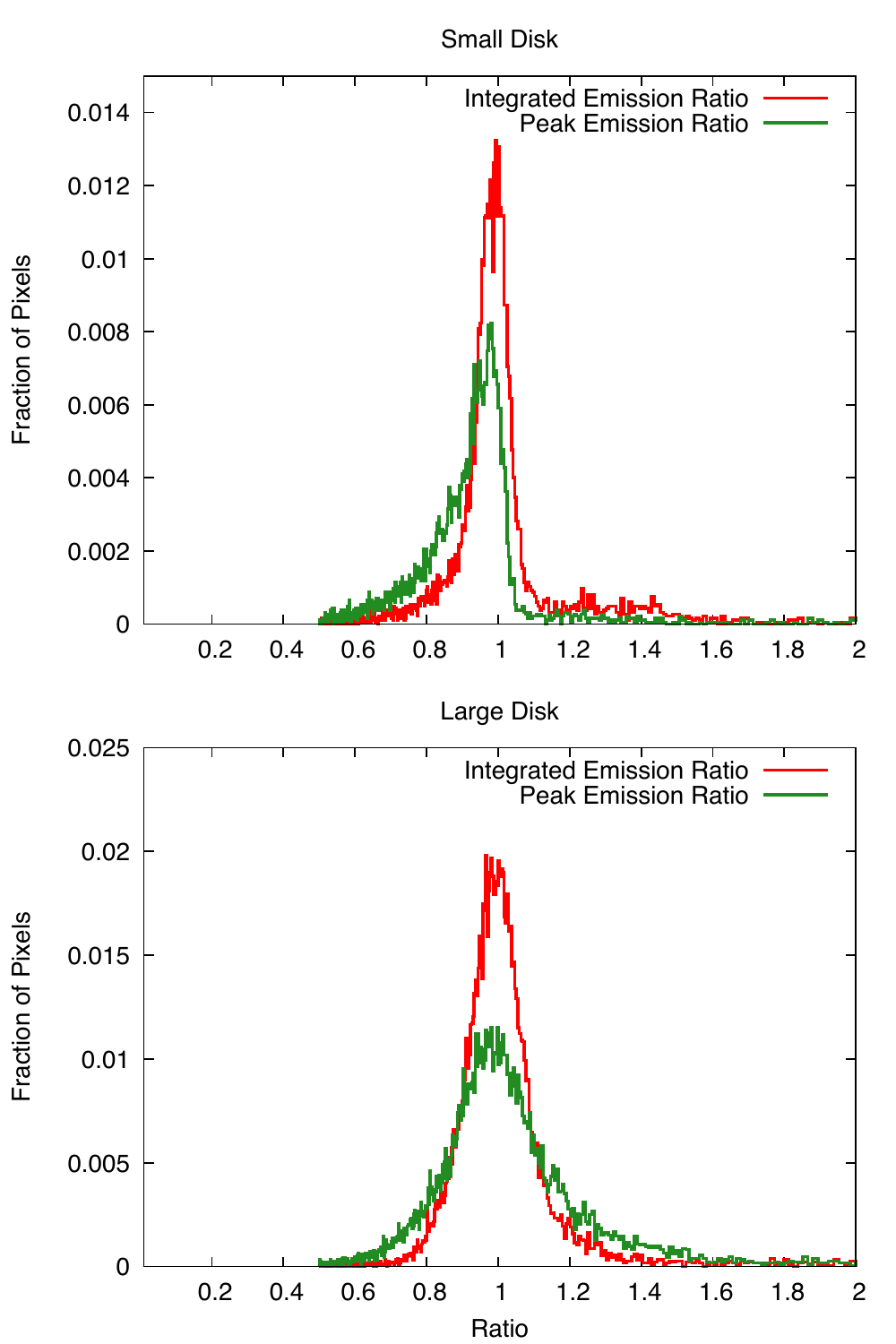}
\caption{\label{Fig:hist3} Top Panel: Ratio of observed model to 'pure' model for the smaller beam size and smaller disk. With noise added, 
the observed model is less accurate for both integrated emission map and peak emission map. Bottom Panel: Ratio of observed model to 
'pure' model for the smaller beam size and larger disk.}
\end{figure}

The top panel of figure~\ref{Fig:hist3} shows a histogram of the ratios of the observed model to the 'pure' model for the disk 
with radius of 100AU and the longer, five hour observation. The main result of the noise being added is that the values for the 
temperature ratio is much more spread out, with most brightness temperatures of the observed model being within 
10-15\% of the original, for any given pixel. Given the larger effect of the 
noise, the differences between using peak emission map and integrated emission map matter less, though the integrated 
emission map is still slightly better. The bottom panel of figure~\ref{Fig:hist3} shows the histogram of the temperature 
ratio, but now for the larger disk. In this case, the two methods give results that are close to the same. It is worth noting that the noise and beam dilution have 
different effects at different regions of the disk. The beam dilution is mostly a problem at the inner and outer regions, where the 
noise is applied evenly across the disk. A result of this is that while the beam dilution is less of a problem as the beam gets 
smaller, the noise is still roughly the same.

This paper has consistently used only the peak emission map, which chooses only the brightest single channel per pixel, and 
the integrated emission map, which integrates all channels with any detectable line emission. However, these two techniques 
represent the extremes, and some observations may work best with a compromise between the two. From the standpoint of noise 
and beam dilution, it is clearly best to average over as many channels as possible, but some of the disks observed in earlier sections 
clearly benefited from the peak emission map. The best results could feasibly come from a technique halfway between the two; using 
multiple channels to reduce noise, but restricting to only the brightest few.

\section{Discussion and Conclusions}

This paper has sought to fully explore the techniques and limitations related to calculating a usable 
brightness temperature from optically thick molecular emission lines in protoplanetary disks. The 
problem of optically thin emission tends to be significant for observations of fainter 
lines. This can come about either because the disk in question is not very massive, or 
because the chosen line is not strong. While $^{12}$CO has many strong rotational transitions, 
the rarer isotopologues are considerably fainter. The problem of continuum subtraction, on the 
other hand is more of an issue for strong lines, and in cases where the continuum emission is 
also very bright. This can be in particularly massive disks, but also in smaller disks which have 
regions of high dust to gas ratio, such as rings, spirals or other asymmetries. 
\citep[See, for example,][]{Boe2017a, Boe2017b, Pere2016}. By introducing peak emission maps as an alternative 
procedure, the problem of optically thin emission can be avoided, and accuracy can be greatly 
improved by proper handling of continuum subtraction.

The conditions that determine whether to use a peak emission map instead of an integrated emission map and whether or 
not to subtract the continuum depend on the conditions and choice of target. While there are cases where no choice can 
yield an accurate brightness temperature, the basic rules are as follows:

\textbullet{} Lighter disks are more susceptible to inclusion of optically thin emission from the line wings. The lighter the disk, 
the better the peak emission map will be relative to the integrated emission map, so long as the peak is optically thick. 

\textbullet{} If both line and continuum are very bright, such as for very massive disks, or for disks with concentrations of dust, 
continuum subtraction can remove substantial amounts of flux. It is important to consider if the line could be significantly 
attenuating the continuum.

\textbullet{} If the beam size is large relative to the size of the channel maps, the peak emission map will lose substantial flux 
to beam dilution, while the integrated emission map will not. When the beam size is small, either technique works well.

These conditions are not hard and fast rules, only guidelines. While it is possible to identify regimes where one technique may 
work better than another, there are not usually strict limits between them. As such, it is necessary that the observer strongly 
consider the following while analyzing data: First, the optical depth of both the line and the dust, when determining whether or not 
to subtract the continuum, and if there are any regions of high dust concentration. Second, the strength of the line, when deciding 
whether to use a peak emission map or integrated emission map. And lastly, the size of the beam and magnitude of the noise.
One caveat to this analysis is that measurements of the gas temperature is affected by systematic uncertainties such as absolute 
flux calibration, which typically ranges between 10-20\%.

The observational pitfalls discussed in this paper have several potentially deep effects to future research into protoplanetary disks 
and planet formation. Firstly, high resolution disk observations have revealed small scale structures in the dust distribution (rings and 
crescents) characterized by surprisingly high dust concentrations. Improper continuum subtraction is a minor effect when continuum 
emission is negligible, but, as shown in Figure~\ref{Fig:MoreDust}, these recently discovered disk structures strongly concentrate dust, 
and therefore emit strong continuum emission which might be optically thick all the way to centimeter wavelengths. See, for example: 
HD142725 \citep{Boe2017a}, MWC758 \citep{Boe2017b}, IRS48 \citep{vdMa2015}, HL Tau \citep{Andr2016}, HD 163296 \citep{Isel2016}. 
Key information about the nature of these structures, which are still 
debated, can be acquired by measuring the amount and temperature of molecular gas using optically thin and optically thick lines, 
respectively. In both cases, the interpretation of the observations requires considerable caution because line and continuum emission 
mutually interact, absorbing each other. As shown in the previous sections, the standard procedure of continuum subtraction can lead to 
a lack of gas emission from regions where the continuum emission is strong. This, in turn, can lead to erroneous estimates of both the 
gas column density and temperature, and consequently, to erroneous physical interpretations. 
For example, the lack of CO emission could be interpreted as the result of the freeze out of molecules on dust grains.
The underestimate of the CO luminosity/temperature might also have an effect on the calculation of the disk turbulence because 
fainter, and therefore colder lines, could results in overestimating the gas turbulence velocity. However, since the turbulence is 
derived in practice by comparing the line intensity with the line broadening \citep[see][]{Flah2015, Flah2017}, it is unclear what 
the combined effect of mutual absorption between gas and dust might be. The derivation of turbulence from CO observations 
will be discussed in a future paper. 

Additionally, ALMA surveys of nearby star forming regions are delivering a large amount of observation of dust and CO emission from nearby 
protoplanetary disks. \citep[See, for example][]{Ansd2017}. The observations could be statistically interpreted in order to estimate disk masses. 
See \cite{Bare2016} or \cite{Mana2016}. Some of the effects shown here (such as continuum subtraction) 
may have an impact on the derivation of gas mass from line intensities. In particular, since 
much gas mass can be hidden in the central and more optically thick regions, continuum subtraction, and, to same extent, 
beam dilution could lead to systematically underestimating the disk mass. This effect will increase with the disk 
mass. Statistical interpretation of line emission and line emission ratios should therefore properly take into account both the 
continuum opacity, and the effect of continuum subtraction on the observations.

Lastly, it is worth mentioning that this paper considered the techniques of determining temperature from single emission lines. In 
many areas of astronomy, an alternate method is often employed to determine temperatures. For a given atom or molecule in 
local thermodynamic equilibrium, the populations of different levels is given by Bolzmann Statistics, so the ratio of two lines will 
be a function only of the temperature. This technique was employed, for example, by \cite{Dart2003} and \cite{Fede2016}. However, this technique is 
not optimal for the applications discussed in this paper for several reasons. Firstly, and most importantly, it is more difficult from 
an observational standpoint because it requires multiple lines to be observed. Attempting to apply this technique to rotational 
modes of CO, for example, would require multiple separate observations, since the multiple transitions rarely fall within any given 
band. Multiple observations are more expensive, and then must be separately calibrated. Since the frequency of the two 
observations varies significantly, the beam size will be different as well. Secondly, the error coming from multiple observations 
will be higher than for a single observation. In general, the errors will add in quadrature, and the already potentially significant 
systematic errors can  become much more problematic. Lastly, from a physical standpoint, the temperature resulting from this 
method is hard to interpret in the case of optically thick emission.Unlike for optically thin observation, where the brightness 
temperature describes a bulk property of the gas along the entire line of sight, each observation of an optically thick line will 
correspond to an emission region at a different depth in the disk, since the opacity depends strongly on the frequency of 
observation. As circumstellar disks have substantial temperature variation, both vertically and radially, these regions will likely have 
different temperature and composition, so a value determined from both may not meaningfully correspond to either. Using line ratios 
to derive a temperature does, however, offer the advantage that the lines observed do not need to be optically thick in order to determine 
a temperature. This makes it potentially useful for extremely light disks, where even $^{12}$CO is optically thin.

\acknowledgments 
Y.B. and A.I. acknowledge support from the NASA Origins of Solar Systems program through award NNX15AB06G. A.I.
acknowledges support from the NSF Grants No. AST-1535809 and AST-1715718. E.W. acknowledges support from the NRAO 
Student Observing Support Grants No. AST-0836064 NSF and AST-1519126

\appendix
\section{Integration of the radiative transfer equation}

In most cases, the synthetic emission from a circumstellar disk must be 
calculated by via numerical integration
of the radiative transfer equation, given a density profile, temperature
profile, and requisite opacities. This process
can be complicated and computationally expensive. So, it is
useful, at least for pedagogical purposes, to consider the case of a face-on, vertically isothermal, axisymmetric
disk, for the simple reason that it presents a case where the
radiative transfer equation becomes analytically solvable.  
In the following derivations, $r$ is the
cylindrical radius and $z$ is the vertical distance from the disk midplane. 

In the relative low density of protoplanetary disks, and for the range
of frequencies considered, scattering is negligible \citep[but, see][]{Kata2015}, the radiative
transfer equation reduces to 
\begin{equation}
\frac{dI_\nu}{ds}=K_\nu(s) \left[ B_{\nu}(T(s)) -  I_\nu(s) \right]
\end{equation}
where $K_\nu(s) = \kappa_\nu(s) \rho(s)$ is the opacity at a given point, and $s$ is 
the spatial coordinate along the line of sight. This assumes that the system is in local 
thermodynamic equilibrium. As discussed in section 2.1, however, LTE is a reasonable 
assumption for the case of CO rotational lines at these temperatures.

In the special case of a face on disk, rays propagate through the
disk aligned with the $z$ axis, which means integration only needs to
be done along this axis, allowing the differential equation to be
solved analytically. Because we have chosen a vertically isothermal
disk structure, $K_{\nu}$ explicitly depends only on r. Inserting
the density profile into equation 5 gives:
\[
\frac{dI}{dz}=\frac{\kappa_{\nu}(r)\cdot\Sigma(r)}{\sqrt{2\pi}h(r)}exp\left(\frac{-z^{2}}{2h^{2}(r)}\right)\left(B_{\nu}(T)-I(z)\right)
\]
This equation is separable:
\[
\frac{dI}{B_{\nu}(T)-I(z)}=\frac{\kappa_{\nu}(r)\cdot\Sigma(r)}{\sqrt{2\pi}h(r)}exp\left(\frac{-z^{2}}{2h^{2}(r)}\right)dz
\]
Integrating both sides yields:
\[
-\ln\left(B_{\nu}(T)-I(z)\right)=\frac{1}{2}\frac{\kappa_{\nu}(r)\cdot\Sigma(r)}{\sqrt{2\pi}h(r)}\sqrt{2\pi}h(r)\cdot erf\left(\frac{z}{\sqrt{2h^{2}(r)}}\right)+C
\]
Which, after cancellation and exponentiation becomes
\[
I(z)=C\cdot\exp\left(-\frac{\kappa_{\nu}(r)\cdot\Sigma(r)}{2}erf\left(\frac{z}{\sqrt{2}h(r)}\right)\right)+B_{\nu}(T)
\]
Using the initial condition, that I=0 at the starting location, $z_{in}$,
we can solve for the constant of integration:
\[
C=-B_{\nu}(T)\exp\left(\frac{\kappa_{\nu}(r)\cdot\Sigma(r)}{2}erf\left(\frac{z_{in}}{\sqrt{2}h(r)}\right)\right)
\]
The final result is quite complicated, but not intractable:
\begin{equation}
I=B_{\nu}(T(r))\left(1-exp\left(\frac{1}{2}\kappa_{\nu}(r)\cdot\Sigma(r)\left(erf\left(\frac{z_{in}}{\sqrt{2}h(r)}\right)-erf\left(\frac{z_{out}}{\sqrt{2}h(r)}\right)\right)\right)\right)
\end{equation}
where $z_{in}$ and $z_{out}$ are the limits of integration along
the line of sight. In most cases, it is practical to define outer
boundaries for the disk, but to good approximation, we can treat the
disk as infinite in vertical extent, and replace the start and end
points with positive and negative infinity. This simplifies the final
solution to the more manageable
\begin{equation}
I=B_{\nu}(T)\left(1-\exp\left(-\kappa_{\nu}(r)\cdot\Sigma(r)\right)\right)
\end{equation}

The optical depth for a given line of sight can be calculated by a
similar process
\begin{equation}
\tau=\int K_{\nu}(s)\rho(s)ds=\kappa_{\nu}(r)\int\frac{\Sigma(r)}{\sqrt{2\pi}h(r)}exp\left(\frac{-z^{2}}{2h^{2}(r)}\right)dz
\end{equation}
\[
\tau=\kappa_{\nu}(r)\frac{\Sigma(r)}{\sqrt{2\pi}h(r)}\cdot\frac{1}{2}\sqrt{2\pi}\cdot erf\left(\frac{z}{\sqrt{2}h(r)}\right)+C
\]
Using the initial condition that the optical depth is zero at the
starting point, the constant of integration can be calculated, and
the final result is:

\begin{equation}
\tau=\frac{\kappa_{\nu}(r)\cdot\Sigma(r)}{2}\left(erf\left(\frac{z_{out}}{\sqrt{2}h(r)}\right)-erf\left(\frac{z_{in}}{\sqrt{2}h(r)}\right)\right)
\end{equation}
This can be simplified using the approximation that $z_{in}$ and
$z_{out}$ are sufficiently large, yielding
\begin{equation}
\tau=\kappa_{\nu}(r)\cdot\Sigma(r)
\end{equation}

The derivation of equation~\ref{Eq:faceOnSolnAdv} follows by a process roughly analagous to the previous cases, except that 
now, the opacity of the emitting material and the opacity of the absorbing material differ. To calculate the attenuated continuum, 
the emitting opacity is only that of the dust, while the absorbing opacity is that of both the dust and the gas. As such, the 
radiative transfer equation reduces to:

\begin{equation}
\frac{dI_\nu}{ds}=\rho(s) \left[ \kappa_{\nu,emit} \cdot B_{\nu}(T(s)) -  \kappa_{\nu,abs} \cdot I_\nu(s) \right]
\end{equation}

Like before, the line of sight is going straight down, and the path is entirely isothermal. Thus, the intensity and temperature depend 
only on the radius. Plugging in the density profile gives

\[
\frac{dI_{\nu}(z)}{ds}=\frac{\Sigma(r)\kappa_{\nu,abs}}{\sqrt{2\pi}h(r)}\cdot exp\left(\frac{-z^{2}}{2h^{2}(r)}\right)\left(\frac{\kappa_{\nu,emit}}{\kappa_{\nu.abs}}B_{\nu}(T(r))-I_{\nu}(z)\right)
\]
The equation can be separated:
\[
\frac{dI_{\nu}(z)}{\frac{\kappa_{\nu,em}}{\kappa_{\nu,abs}}B_{\nu}(T(r))-I_{\nu}(z)}=\frac{\Sigma(r)\kappa_{\nu,abs}}{\sqrt{2\pi}h(r)}\cdot exp\left(\frac{-z^{2}}{2h^{2}(r)}\right)dz
\]
Integrating both sides yields:
\[
-ln\left(\frac{\kappa_{\nu,em}}{\kappa_{\nu,abs}}B_{\nu}(T(r))-I_{\nu}(z)\right)=\frac{\Sigma(r)}{2}\kappa_{\nu,abs}\cdot erf\left(\frac{z}{\sqrt{2}h(r)}\right)+C
\]
where C is a constant of integration. Thus:
\[
I_{\nu}(z)=\frac{\kappa_{\nu,em}}{\kappa_{\nu,abs}}B_{\nu}(T(r))-C\cdot\exp\left(-\frac{\Sigma(r)}{2}\kappa_{\nu,abs}\cdot erf\left(\frac{z}{\sqrt{2}h(r)}\right)\right)
\]
We can use the initial condition that $I_{\nu}(z_{in})=0$ to solve for the constant of integration:
\[
C=\frac{\kappa_{\nu,em}}{\kappa_{\nu,abs}}B_{\nu}(T(r))\cdot exp\left(\frac{\Sigma(r)}{2}\kappa_{\nu,abs}\cdot erf\left(\frac{z_{in}}{\sqrt{2}h(r)}\right)\right)
\]
Plugging this value in and simplifying gives:
\[
I_{\nu}(z)=\frac{\kappa_{\nu,em}}{\kappa_{\nu,abs}}B_{\nu}(T(r))\left(1-exp\left(\frac{\Sigma(r)}{2}\kappa_{\nu,abs}\left(erf\left(\frac{z_{in}}{\sqrt{2}h(r)}\right)-erf\left(\frac{z}{\sqrt{2}h(r)}\right)\right)\right)\right)
\]
And, using the simplifying assumption that the ray begins at minus infinity and goes to infinity, we recover equation~\ref{Eq:faceOnSolnAdv}:
\[
I_{\nu}(z)=\frac{\kappa_{\nu,em}}{\kappa_{\nu,abs}}B_{\nu}(T(r))\left(1-exp\left(-\Sigma(r)\cdot\kappa_{\nu,abs}\right)\right)
\]

\section{CO Opacity Model}

The opacity of a given point in space is the sum of the continuum opacity and CO opacity.
Continuum opacity must be calculated using Mie Theory, for which several
programs already exist. CO opacity is simpler, and can be derived
relatively easily from first principles. Broadly speaking, for a transition
between states i and j, energy $h\nu_{0}$ is absorbed from the radiation
$I_{\nu}$, where $\nu_{0}$ is the rest energy of the transition.
The relation is given by
\begin{equation}
\kappa_{ij}I_{\nu}d\nu d\Omega=h\nu_{0}\left(n_{i}\phi_{\nu}B_{ij}I_{\nu}-n_{j}\psi_{\nu}B_{ji}I_{\nu}\right)\frac{d\nu d\Omega}{4\pi}
\end{equation}
where $\kappa_{ij}$ is the mass absorption coefficient, $\phi_{\nu}$
and $\psi_{\nu}$ are the absorption and emission profiles, $B_{ji}$
and $B_{ij}$ are the Einstein Coefficients for stimulated emission
and photo absorption. The shape of the profiles can vary, but must
always be normalized to unity. Solving for the mass absorption coefficient
gives
\begin{equation}
\kappa_{ij}=\frac{h\nu_{0}n_{i}\phi_{\nu}B_{ij}}{4\pi}\left(1-\frac{n_{j}\psi_{\nu}B_{ji}}{n_{i}\phi_{\nu}B_{ij}}\right)
\end{equation}
This can be simplified using one of the Einstein Relations: 
\[
g_{i}B_{ij}=g_{j}B_{ji}
\]
and, that under conditions of Local Thermodynamic Equilibrium, Boltzmann
Statistics relates the populations through
\[
\frac{n_{j}}{n_{i}}=\frac{g_{j}}{g_{i}}\exp\left(-\frac{h\nu_{0}}{kT}\right)
\]
to give
\begin{equation}
\kappa_{ij}=\frac{h\nu_{0}n_{i}\phi_{\nu}B_{ij}}{4\pi}\left(1-\exp\left(-\frac{h\nu_{0}}{kT}\right)\right)
\end{equation}
Here we have also used the assumption of complete redistribution,
which means that the absorption and emission profiles are the same.
This expression is most frequently expressed using the Einstein A
value, related through the Einstein relation
\[
A_{ji}=\frac{2h\nu}{c^{2}}Bji
\]
to obtain
\begin{equation}
\kappa_{ij}=\frac{c^{2}}{8\pi\nu_{0}^{2}}\cdot\frac{g_{j}}{g_{i}}n_{i}\phi_{\nu}A_{ji}\left(1-\exp\left(-\frac{h\nu_{0}}{kT}\right)\right)
\end{equation}

The number of molecules of CO currently in state i, $n_{i}$, is given
by the Boltzmann equations:
\[
n_{i}=n_{CO}\cdot\frac{g_{i}\exp\left(-\frac{E_{i}}{kT}\right)}{Z(T)}
\]
where $n_{CO}$ is the number of CO particles per gram of gas, $E_{i}$
is the energy of the level, and Z(T) is the partition function. For
rotational transitions, the partition function can be expressed as
\[
Z(T)=\sum_{i=0}^{\infty}\left(2i+1\right)\exp\left(\frac{-E_{i}}{kT}\right)
\]
The partition function can also be evaluated quickly and with high
accuracy via power series expansion:
\begin{equation}
Z(T)=\frac{kT}{hB_{0}}+\frac{1}{3}+\frac{1}{15}\left(\frac{hB_{0}}{kT}\right)+\frac{4}{315}\left(\frac{hB_{0}}{kT}\right)^{2}+\frac{1}{315}\left(\frac{hB_{0}}{kT}\right)^{3}+\ldots
\end{equation}
where $B_{0}$ is the rigid rotor rotation constant for the molecule.
For more detail, see \cite{MaSh2015}. $\phi_{\nu}$ is the natural
line profile, is given by
\begin{equation}
\phi_{\nu}=\frac{c}{\nu_{0}}\cdot\frac{1}{\Delta V\sqrt{\pi}}\exp\left(-\frac{\Delta^{2}\nu}{\Delta^{2}V}\right)
\end{equation}
with $\Delta\nu$ being the line of sight velocity of the gas. This
is a gaussian profile, typical of thermalized gas, but is generalized
to include both thermal broadening and possible turbulent broadening,
both included in 
\begin{equation}
\Delta V=\sqrt{\frac{2kT}{m_{CO}}+V_{turb}^{2}}
\end{equation}
While the equations above are fully general, the calculations done
in this paper assume there is no turbulence in the gas. Though other
sources of broadening exist, none are particularly relevant under
these conditions. Furthermore, in several of the cases considered in this paper,
the disk is viewed face on, the line of sight velocity resulting from
Keplerian rotation is zero.

The Einstein A value can, in principle, be calculated from theoretical
models. Mihalas gives the Einstein A value for dipole emission as
\begin{equation}
A_{ji}=\frac{64\pi^{4}\nu^{3}}{3hc^{3}}\left|\mu\right|^{2}
\end{equation}
where $\left|\mu\right|^{2}$ is the squared dipole matrix element
summed over degenerate states, and can in this case be expressed as
\[
\left|\mu_{ji}\right|^{2}=\frac{j+1}{2j+1}\mu^{2}
\]
with $\mu$ being the permanent dipole moment of the molecule. (See 
\cite{HuMi2015}, or \cite{Spit1978} for further details.) It is worth noting, however, that the
theoretical values do not agree particularly well with experimentally
measured values, such as those from the Leiden LAMDA database. This
paper, therefore, uses the measured values instead.

\section{Ray Tracing}

While Section A described an analytical solution to the radiative transfer equation, it was predicated by several simplifying assumptions. 
The disk structure was taken to be vertically gaussian, the temperature vertically isothermal, and most importantly, the line of sight was 
also vertical. While useful as a proof of concept, this model is far too simple to be sufficient. However, adding any further details 
complicates the problem enough that no analytical solution is viable. Toward that end, this paper uses a new, numerical ray tracing code 
to analyze more realistic disk models.

The code is written entirely in C++, using the new language features and libraries introduced with the C++11 standard. Unlike other 
commonly used ray tracing codes, (for example, RADMC 3D), it does not use a gridded disk structure. Gridded models work by dividing 
the disk in radius, zenith, and azimuth bins, each with a separate density, temperature, and composition. Ray tracing is then done by 
casting a ray through the disk and summing the contribution and attenuation due to each grid cell the ray passes through. This approach 
has several advantages. First, the output of most radiative transfer codes is in the same gridded form, because radiative transfer of this 
form must be done via Monte Carlo methods. Second, the calculation of the emission is very simple. The primary disadvantage is that 
the gridded approach is slowed down by the overhead of calculating the correct grid cells to include. Determining which cell to enter 
requires calculating the intersections with all nearby edges, which is nontrivial, and leads to numerical instabilities at the corners. 
Additionally, while easy to implement, this is a first order method of solving the radiative transfer equation. This makes it well behaved, 
but extremely slow.

\begin{figure}[!t]
\centering
\includegraphics[width=\linewidth]{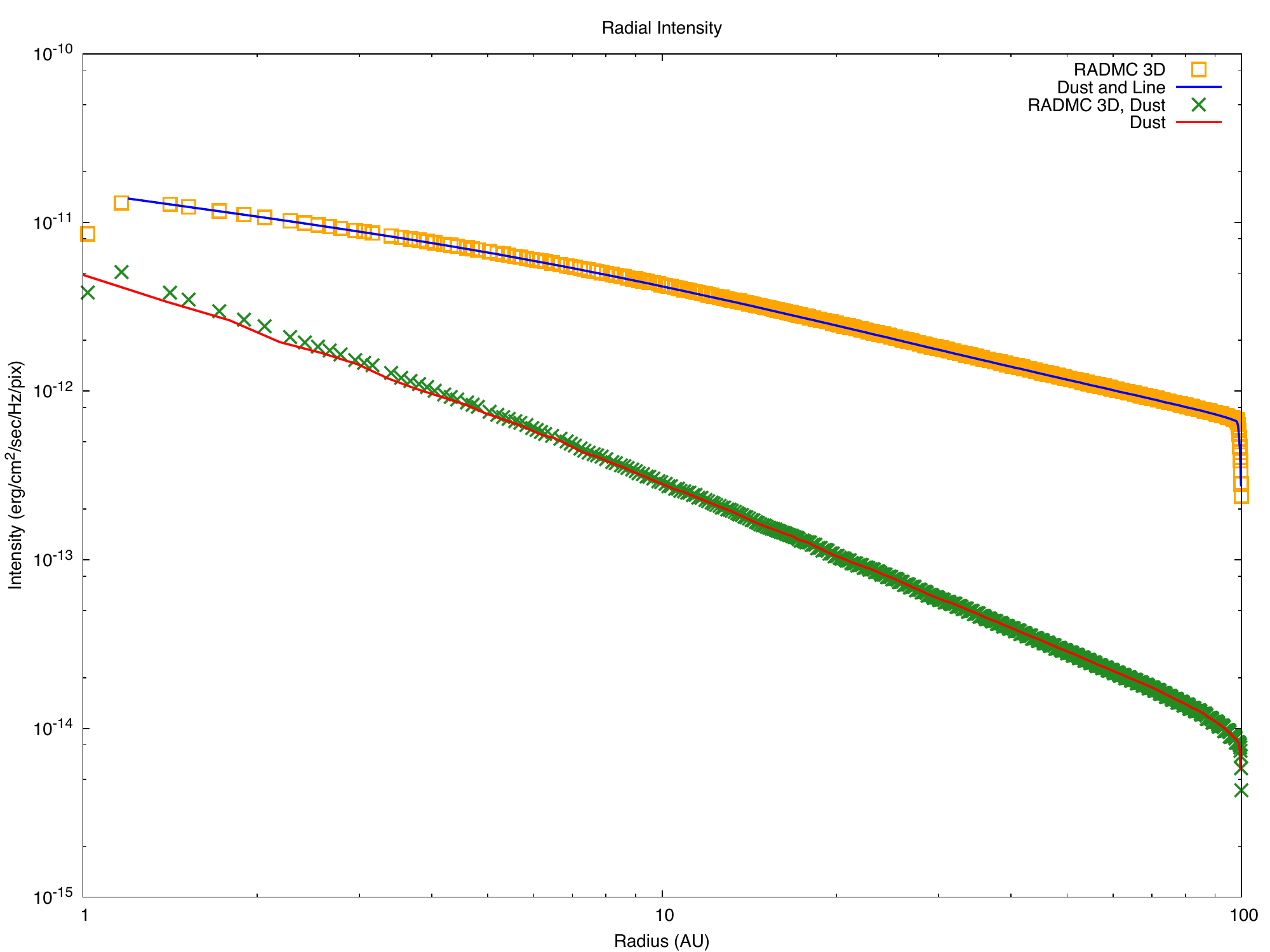}
\caption{\label{Fig:RayTraceComp} Agreement between outputs from the code used in this paper and RADMC 3D. The simulation in this 
case is a face on disk with a mass of $2\times10^{-2} M_{odot}$, with an outer radius of 100 AU. Two sets of data are presented: the blue 
and orange correspond to the models including both continuum and line emission, where the red and green correspond to the models 
that contain only continuum emission. The lines are the output of the code used in this paper, where the points are the results from RADMC 
3D.}
\end{figure}

The code used in this paper opts to instead use a non-gridded approach. This means that input files, such as the temperature, must be 
interpolated on, but this doesn't add much computational time. Considerable time is saved by removing all of the geometric calculations. 
Furthermore, the differential equation doesn't need to be solved with a first-order method. In theory, this means that a higher order adaptive 
algorithm (such as the commonly used combined 4th-5th order Runge Kutta scheme, and its higher order replacements), but in practice 
the radiative transfer equation is difficult to numerically solve. Almost all emission
is generated from a region around the $\tau=1$ surface: In front of this region, the density is low enough that there is minimal emission,
and behind it, the opacity is high enough that any emission is attenuated away. Simple numerical integration schemes which approximate 
the solutions to differential equations as smooth and low order polynomials have difficulty with such a sharp solution. Additionally, adaptive 
step size algorithms increase their steps in larger and larger amounts as they move through the initial region, and then usually completely 
overstep the emission region, giving a completely wrong final result. In the future, it may be possible to work around these problems, but for 
the current time, we are restricted to the implicit 
trapezoid rule instead. There are several reasons for this: Most importantly, the radiative transfer equation without scattering, as 
written in eq. 5 is a stiff equation. An equation is stiff if its exact solution has a term which decays as an exponential, but has derivatives 
which are larger than the terms they correspond to. Stiff equations cannot be solved with explicit numerical schemes, because these 
algorithms will diverge. Instead, implicit techniques must be used. Second, the trapezoid rule is the highest order integration scheme which 
is absolutely convergent. This means that it will converge to the correct solution regardless of the step size. Though it does not converge as 
quickly as a higher order method would, it is guaranteed to converge to the correct solution, unlike other methods.

In practice, the use of a non-gridded disk with the implicit trapezoid rule offers a speed boost of roughly twice that of other methods tested 
(mainly RADMC 3D). This boost in speed is made possible possible primarily by the fact that this code is highly optimized for this specific 
problem, where other codes like RADMC 3D are designed to be much more general, and can handle other geometries, or even include 
scattering. In order to further increase the efficiency of simulations, the code is multithreaded, which allows the speed to scale linearly with 
the number of CPU cores allotted.

The image is constructed of an array of pixels, each corresponding to a line of sight cast out normal to the image surface. All imaging
is done under the assumption that the disk is so distant that the small angle approximation is valid and lines of sight are approximately
parallel, a reasonable assumption, given that even the closes disks are still dozens of parsecs away. Imaging is multithreaded so that 
multiple rays can be done in parallel.

Results from disk imaging have been tested for accuracy against both the analytical solution described in Appendix A, as well as other
codes written for similar processes, such as RADMC 3D and RADEX. Results match the analytical solution almost exactly, and are close 
to matching output from both other ray tracing codes. In terms of speed, this approach is considerably faster than other available ray tracing 
codes, both for single processor comparisons and through being multithreaded. Figure~\ref{Fig:RayTraceComp} shows the matching in 
results between the code used for this paper and output for an identical disk model from RADMC 3D. The model shown is a disk of mass 
$2\times10^{-2} M_{odot}$, with an outer radius of 100 AU. Two sets of data are presented: the blue 
and orange correspond to the models including both continuum and line emission, where the red and green correspond to the models 
that contain only continuum emission. The lines are the output of the code used in this paper, where the points are the results from RADMC 
3D. The agreement is quite close, despite the fundamentally different integration schemes.

\end{document}